\documentclass[prl,twocolumn,amsmath,amssymb,english]{revtex4-1}

\usepackage[dvipdfm]{graphicx}
\usepackage{natbib}
\usepackage{subfigure}
\usepackage{tabularx}
\usepackage{epsfig}
\usepackage{longtable}
\usepackage{amsfonts}
\usepackage{rotating}
\usepackage{subfigure}
\usepackage{amsmath}
\usepackage{hyperref}
\usepackage{babel}

\usepackage{color}

\newcommand{\bea}{\begin{eqnarray}}
\newcommand{\eea}{\end{eqnarray}}
\newcommand{\la}{\label}
\newcommand{\be}{\begin{equation}}
\newcommand{\ee}{\end{equation}}



\begin{document}

\title{Critical integer quantum Hall topology and the integrable Maryland model as a topological quantum critical point}
 \author{Sriram Ganeshan}
\author{K. Kechedzhi}
\author{S. Das Sarma}
\affiliation{Condensed Matter Theory Center and Joint Quantum Institute, Department of Physics, University of Maryland, College Park, MD 20742, USA}

\date{\today}

\begin{abstract}

One dimensional tight binding models such as Aubry-Andre-Harper (AAH) model (with onsite cosine potential) and the integrable Maryland model (with onsite tangent potential) have been the subject of extensive theoretical research in localization studies.  AAH can be directly mapped onto the two dimensional Hofstadter model which manifests the integer quantum Hall topology on a lattice. However, no such connection has been made for the Maryland model (MM). In this work, we describe a generalized model that contains AAH and MM as the limiting cases with the MM lying precisely at a topological quantum phase transition (TQPT) point. A remarkable feature of this critical point is that the 1D MM retains well defined energy gaps whereas the equivalent 2D model becomes gapless, signifying the 2D nature of the TQPT. 

\end{abstract}

\maketitle


Integer Quantum Hall Effect (IQH) is a canonical example of a gapped bulk topological phase with no generic symmetry protection. IQH can be captured by the 2D Hofstadter model~ \cite{hofstadter,laughlin,tknn,Avron,hatsugai,Ponomarenko_2013,Dean_2013,Hunt_2013,ketterle,blochprl}, a 2D lattice tight binding model with non-zero flux per unit cell. Hofstadter model can be mapped onto the 1D Aubry-Andre-Harper \cite{harper55,AA} (AAH) model, a 1D tight binding chain with onsite cosine potential. 
Aubry and Andr\'e~\cite{AA} identified a localization transition in the AAH model with modulation incommensurate with the lattice (corresponding to an irrational value of flux). This result led to an extensive theoretical investigation of the AAH model in the context of localization studies~\cite{AA,sankarprl88, thoulessprl88, sankarprb90, biddlepra09, biddleprl10}. Recent experimental developments in photonic crystals~\cite{lahini2009, kraus1, kraus3} and ultracold atoms~\cite{wiersma, roati08,modugno} have realized this localization phenomena in 1D quasiperiodic AAH lattices.

A completely different example of a 1D tight binding model with an onsite tangent modulation is presented by the 1D `Maryland model'.  The `Maryland Model' (MM) was proposed and solved exactly by Grempel et. al.~\cite{grempel, prange,fishman}. MM has one-to-one correspondence with the quantum kicked rotor problem which has been experimentally realized in ultra cold atoms~\cite{garreau}, and has been extensively studied~\cite{berry,simon}. We discover in the current work a completely unexpected deep mathematical connection between MM and IQH, which has remained unappreciated in the literature. In addition, we show that MM presents an intriguing example of a topological quantum phase transition (TQPT).

Maryland model with period of onsite potential incommensurate with lattice spacing presents an example of a 1D quasicrystal (QC) for which a special `quasiperiodic' transatlion symmetry was recently identified~\cite{kraus1}.  A family of 1D QCs taken together (generalized AAH,  Fibonacci~\cite{fibonacci}) ~\cite{kraus2,kraus3} has been topologically classified with an equivalent IQH topology in 2D. This classification was identified by connecting different models of QCs with the same topological invariant corresponding to the real space 2D lattice with a flux~\cite{kraus2}. An argument was made~\cite{kraus1} and subsequently debated~\cite{brouwer,krauscomment} that this quasiperiodic translation symmetry allows one to associate 2D IQH invariants to each 1D member of the family of QCs~\cite{kraus1}. The fact that the MM belongs to this 1D quasicrystal symmetry class and was not associated with the IQH topology calls for an investigation of this model from a fresh perspective. We base our arguments only on the well established connection between families of 1D tight binding models with periodic modulation and 2D IQH topology~\cite{harper55,hofstadter,AA}.

In this letter, we take a fresh approach in understanding the relationship between the IQH topology and the Maryland model. We construct a  family of 1D tight binding models parameterized by a phase with a general onsite modulation potential that contains AAH and MM as limiting cases. 
We construct the equivalent real space 2D lattice model by taking an inverse Fourier transform with respect to this phase parameter. We analyze the energy spectrum of the general 1D model as a function of the phase parameter. We identify the topological invariants for this general model using the theory of electric polarization~\cite{vanderbilt,restaprb} which provides a natural framework to study IQH invariants. Based on this analysis we explicitly show that the Maryland model sits at the critical point of a quantum phase transition to the topologically trivial state. The criticality of the Maryland model allows us to associate topological invariants to it in a purely mathematical sense using the limiting procedure along the deformation path in the parameter space. We show that even though the 1D gaps are preserved throughout the deformation from AAH to MM, the energy gaps in the equivalent 2D model close at the TQPT, as required by general considerations. We discuss the consequences of this result for the topological classification of 1D QC families~\cite{kraus1,brouwer,krauscomment}. 

We consider a 1D tight binding chain of size $N$ with an onsite potential modulation $V_n(\alpha,\varphi)$,
\begin{gather}
H(\varphi,\alpha)=-\sum_{n=1}^{N-1} t (c^{\dagger}_{n+1}c_n+c^{\dagger}_n c_{n+1})-\sum_{n=1}^{N} V_n(\alpha,\varphi)c^{\dagger}_{n}c_n,\la{eq:Hamiltonian}\\
V_{n}(\alpha,\varphi)=2\text{\ensuremath{\lambda}}\left(\frac{\cos\left(2\pi n b+\varphi-\alpha\frac{\pi}{2}\right)}{1+\alpha\ \cos(2\pi n b+\varphi)}\right),\nonumber
\end{gather}
%
where $c^\dagger_n$ and $c_n$ are 
creation and annihilation operators on the site $n=1$, $2$, $\ldots$, $N$, and $t$ is the amplitude of the nearest neighbour hopping. 
The onsite potential, $V_{n}(\alpha,\varphi)$, is characterized by the strength $\lambda$, period $1/b$ and the phase parameter $\varphi$. The parameter $\alpha$  interpolates between the limiting cases AAH ($\alpha=0$) and MM ($\alpha=\pm1$), 
\bea
V_{n}(\alpha,\varphi)
&=&2\lambda\begin{cases}
\cos(2\pi n b+\varphi) & for\ \alpha\rightarrow 0\\
\\
\left(\tan\left(\frac{2\pi n b+\varphi}{2}\right)\right)^{\alpha} & for\ \alpha \rightarrow \pm1
\end{cases}
\eea
This general onsite potential is a smooth function of $\alpha$ in the open interval $\alpha \in (-1,1)$. $V_{n}(\alpha,\varphi)$ has singularities at $\alpha=\pm1$ corresponding to the integrable MM which we approach asymptotically in a limiting sense and we define TQPT in terms of these singularities.  $V_{n}(\alpha,\varphi)$ is a specific example of a generic $2\pi$ periodic onsite potential  $\mathcal{F}[2\pi n b+\varphi]$, where  $\mathcal{F}(z)$ is an analytic function everywhere except in the limit of singular MM where it acquires isolated poles.  
 %
%
\begin{figure}[htb!]
  \centering
\includegraphics[width=8cm,height=5cm]{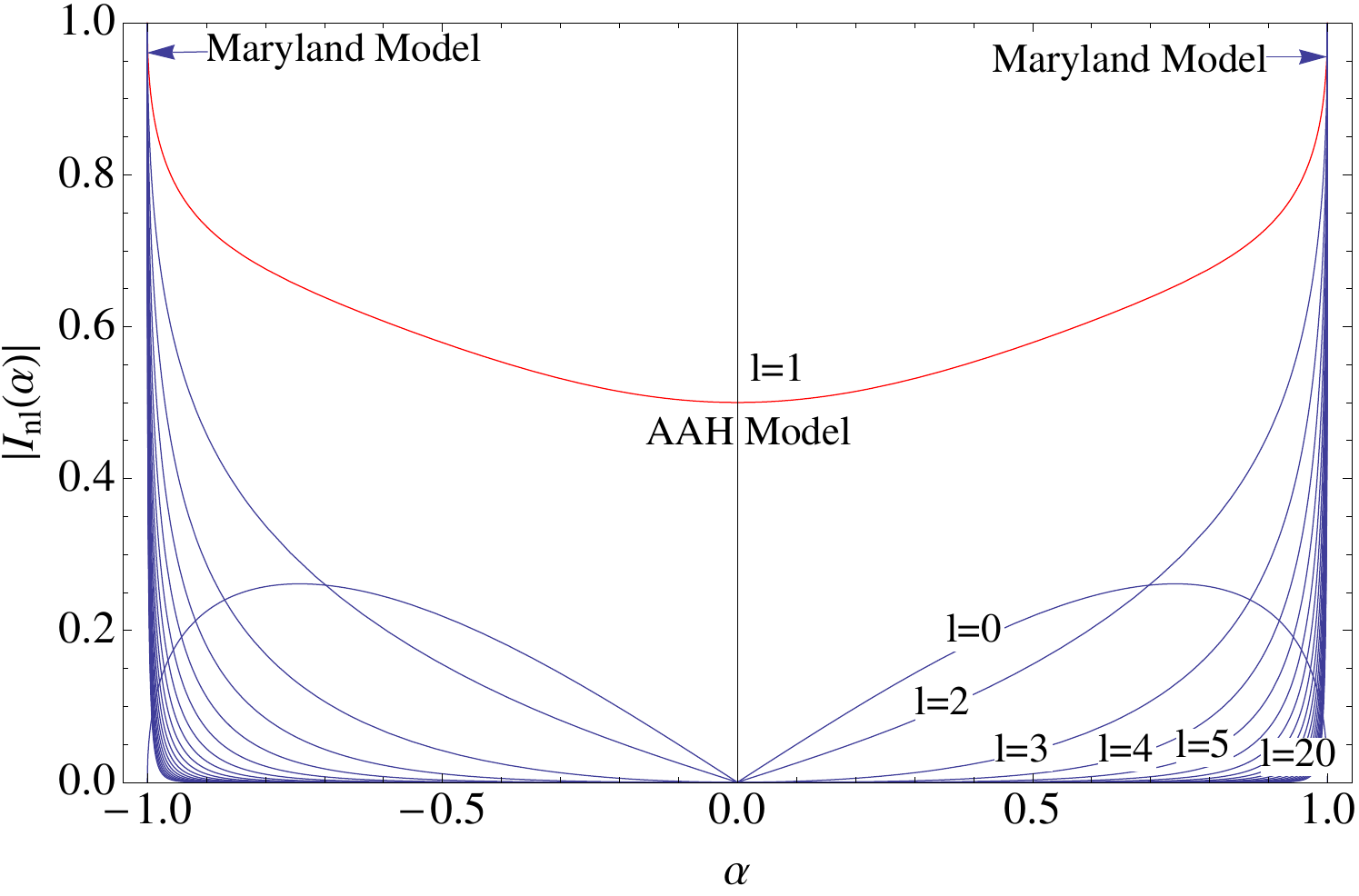}
  \caption{ $|I_{nl}(\alpha)|$, as a function of $\alpha$ for different hopping range $l$. As $\alpha\rightarrow 0$ (AAH model) only the nearest neighbor hopping term is non-zero ($l=1$, shown in red). Long-range hopping amplitudes increase with $\alpha$ and in the limit $\alpha\rightarrow \pm1$ (Maryland model) the hopping amplitudes of all ranges are equal to unity ($|I_{nl}(\alpha)|=1$). }
  \label{2dlattice}
\end{figure}

\textit{2D ancestor:-} 
Taking an inverse Fourier transform with respect to $\varphi$ results in a real space lattice which is the 2D Hofstadter model with a flux $b$ per unit cell. 
 The same idea applies to the Hamiltonian in Eq.~(\ref{eq:Hamiltonian}),
\begin{gather}
H_{2D}(\alpha)=\int_{-\pi}^{\pi}d\varphi H(\varphi,\alpha)e^{im\varphi},\nonumber\\
c_{n}\equiv c_{n}(\varphi)=\frac{1}{\sqrt{2\pi}}\sum_{m}e^{i m\varphi}c_{n,m},
\la{ft}
\end{gather}
where $n,m=1,..,N$. 
The resulting real space 2D Hamiltonian equivalent to $H(\varphi,\alpha)$ (up to a constant energy shift) reads,
\bea
H_{2D}(\alpha)&=&\sum_{m,n}[t(c_{n,m}^{\dagger}c_{n+1,m}+c_{n+1,m}^{\dagger}c_{n,m})\nonumber\\
&+&2\lambda\sum^{\infty}_{l=0}(I_{nl}(\alpha)c_{n,m}^{\dagger}c_{n,m-l}+h.c.)], \label{H_{2D}alpha}
\eea
where
\bea
I_{nl}(\alpha)&=&e^{-il(2\pi nb+\alpha\frac{\pi}{2})}\bigg[\frac{e^{i \pi  \alpha }}{\alpha }\delta _{l,0}+\left(-1+\sqrt{1-\alpha ^2}\right)^{l-1}\nonumber\\
&\times&\frac{ \left(2-\alpha ^2(1-e^{i \pi  \alpha } )-2 \sqrt{1-\alpha ^2}\right)}{2 \alpha ^{l+1} \sqrt{1-\alpha ^2} }\bigg].
\la{hop}
\eea
describes the hopping amplitude from site $m-l$ to $m$, i.e. a hopping of range $l$ ($l=0$ term is the constant shift in the onsite energy). Note that $I_{nl}(\alpha)$ in Eq.~(\ref{hop}) is defined in a limiting sense at the special points $\alpha=0,\pm1$ (AAH and MM)~\cite{supp}.
%
%
%
Fig.~\ref{2dlattice} plots the absolute value of the hopping amplitude $|I_{nl}(\alpha)|$ for different values of the hopping range $l$ as a function of $\alpha$. In the limiting case of $\alpha\rightarrow 0$ (AAH) only the $l=1$ term survives, $I_{n1}(0)=1/2$, which corresponds to the nearest neighbor hopping of the Hofstadter model. 
%
As $\alpha$ increases, hopping terms of longer range $l$ in the $m$-direction acquire non-vanishing amplitudes. In the limiting case of $\alpha\rightarrow\pm1$ (MM) the dual 2D lattice acquires long range hopping terms of arbitrarily large $l$ in the $m$-direction all of equal unit amplitude, see Eq.~(\ref{hop}). 
This arbitrarily long range hopping singularity is indicative of a quantum phase transition occurring at the critical points $\alpha=\pm1$. To elucidate the physical nature of these $\alpha=\pm1$ critical points further, we analyze the band structure and the topological invariants of Eq.~(\ref{eq:Hamiltonian}) as a function of $\alpha$. 
\begin{figure}[htb!]
  \centering
 \includegraphics[width=4cm,height=3cm]{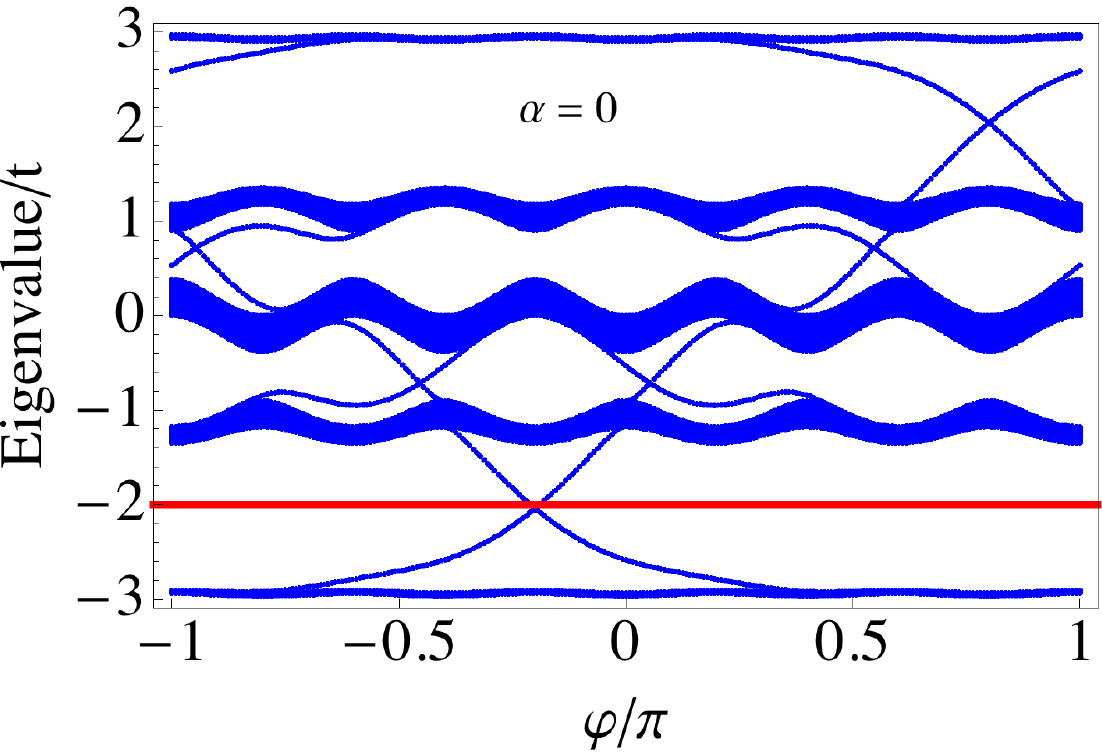}\hspace{0.1cm}\includegraphics[width=4.0cm,height=3cm]{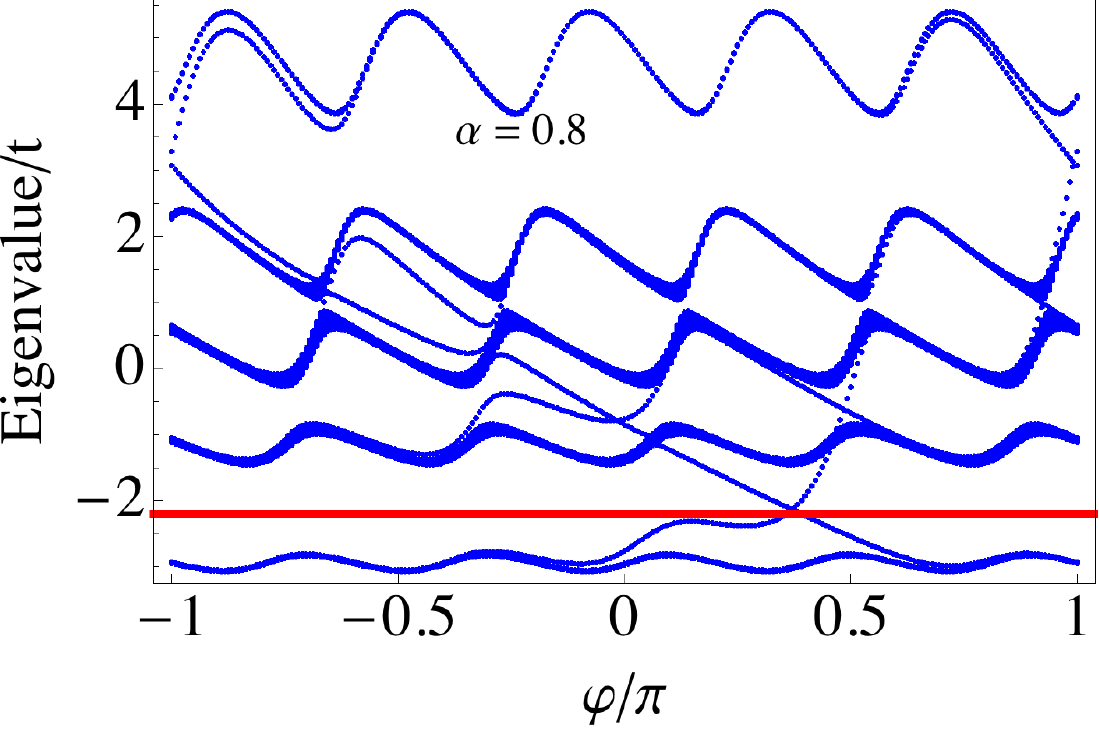}\\
\hspace{-0.2cm}
\includegraphics[width=4.0cm,height=3cm]{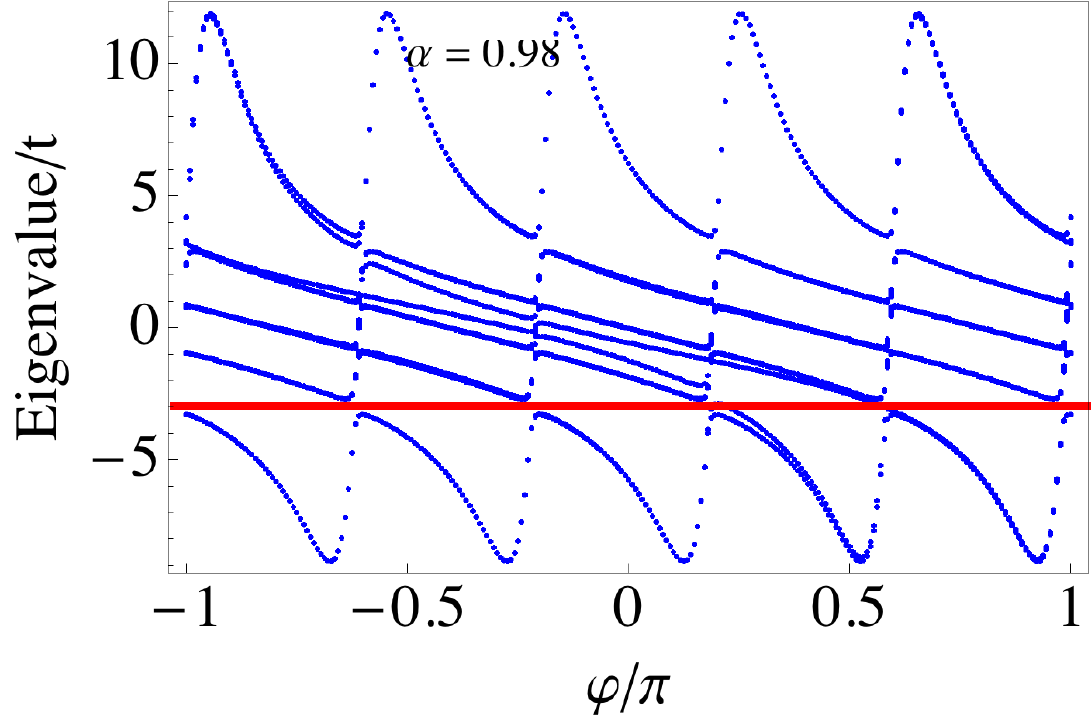}\hspace{0.1cm}\includegraphics[width=4.0cm,height=3cm]{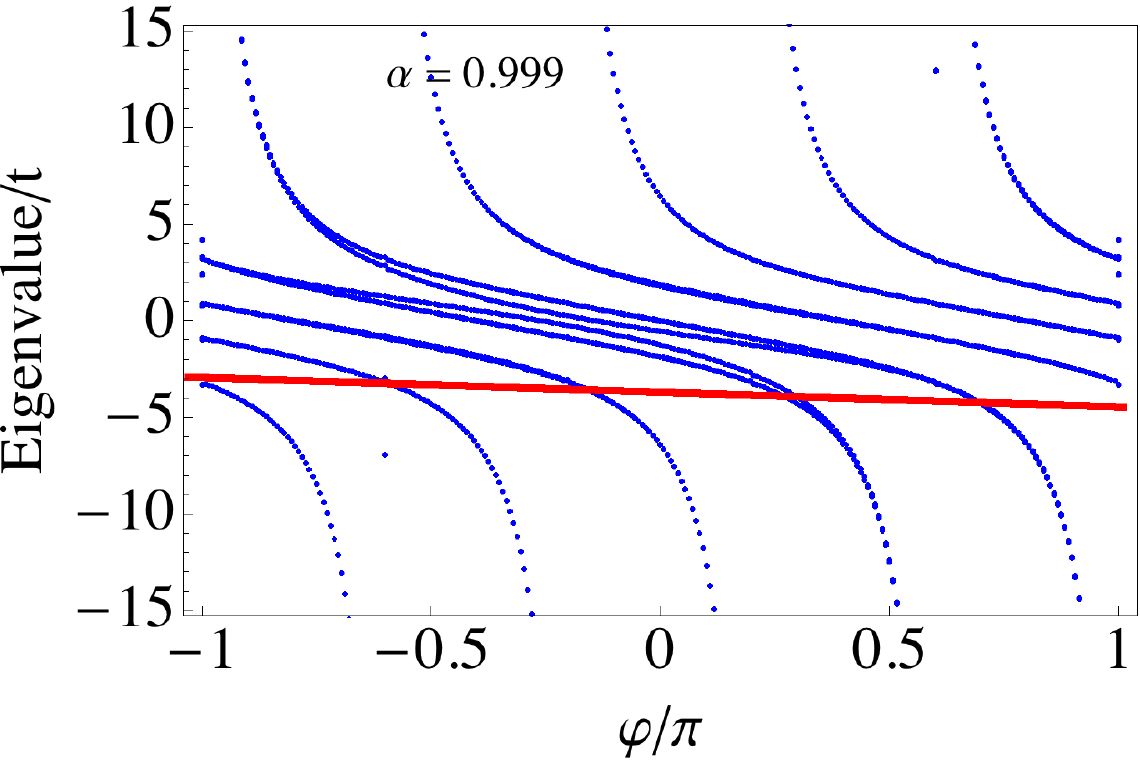}
    \caption{The energy spectrum plotted for $N=200$ sites, $b=1/5$ and $\lambda=1$ for $\alpha=0$ (AAH model),  $\alpha=0.8$,  $\alpha=0.98$ and  $\alpha=1.0$ (Maryland model). The red line separates empty and filled states in the spectrum. Instead of fixing the Fermi level in our numerics we fix the number of particles per site since the latter can be fixed thorughout the deformation  driven by the parameter $\alpha$ and even at $\alpha=1$.}
  \la{energy}
\end{figure}
\vspace{-0.015cm}

\textit{Band structure:-} 
We impose open boundary conditions on the 1D tight binding Hamiltonian, $H(\varphi,\alpha)$, in Eq.~(\ref{eq:Hamiltonian}) and numerically diagonalize it for the system size of $N=200$ sites. It is instructive to plot the resulting energy bands as a function of the phase parameter $\varphi$, which captures the 2D band structure in the hybrid space $(n,\varphi)$. We start with the case of a commensurate modulation by setting $b=1/5$ and $\lambda=1$. Fig.~\ref{energy} shows the resulting band structure as a function of the phase parameter $\varphi$ for four different values of $\alpha$. The case of AAH ($\alpha=0$), top left panel of Figure~\ref{energy}, demonstrates a well defined set of gaps reflecting the robust integer quantum Hall topology of the 2D Hofstadter model. For $\alpha=0.8$ and $\alpha=0.98$, the band gaps gradually decrease. All band gaps close (scale to zero with the system size) precisely at the critical point $\alpha=1$ (as explicitly shown using the exact spectrum \cite{supp,watson}). 
The gapless nature of the 2D spectrum for the MM case ($\alpha=\pm1$) in the hybrid space $(n,\varphi)$ is explicitly confirmed using the exact analytical expression for the MM spectrum with commensurate modulation~\cite{watson}. 

%
Closing of the spectral gaps coincides with the hopping range divergence in the 2D lattice and indicates a TQPT in the system as $\alpha \rightarrow \pm1$ (i.e. at the MM point). The interesting aspect of MM is that the 2D spectrum is gapless in the reciprocal space $(k_n,\varphi)$ whereas the 1D spectrum has well defined gaps for each value of $\varphi$. Here $k_n$ is the Fourier image of the site index $n$. The fact that the 1D MM spectrum has gaps whereas the corresponding dual 2D spectrum is gapless makes perfect sense since the non-trivial TQPT can only exist in the 2D space. The scale invariance of the system at the transition point can also be explicityly demonstrated~\cite{supp}.

%
 \begin{figure}[htb!]
  \centering
\includegraphics[width=4cm,height=3cm]{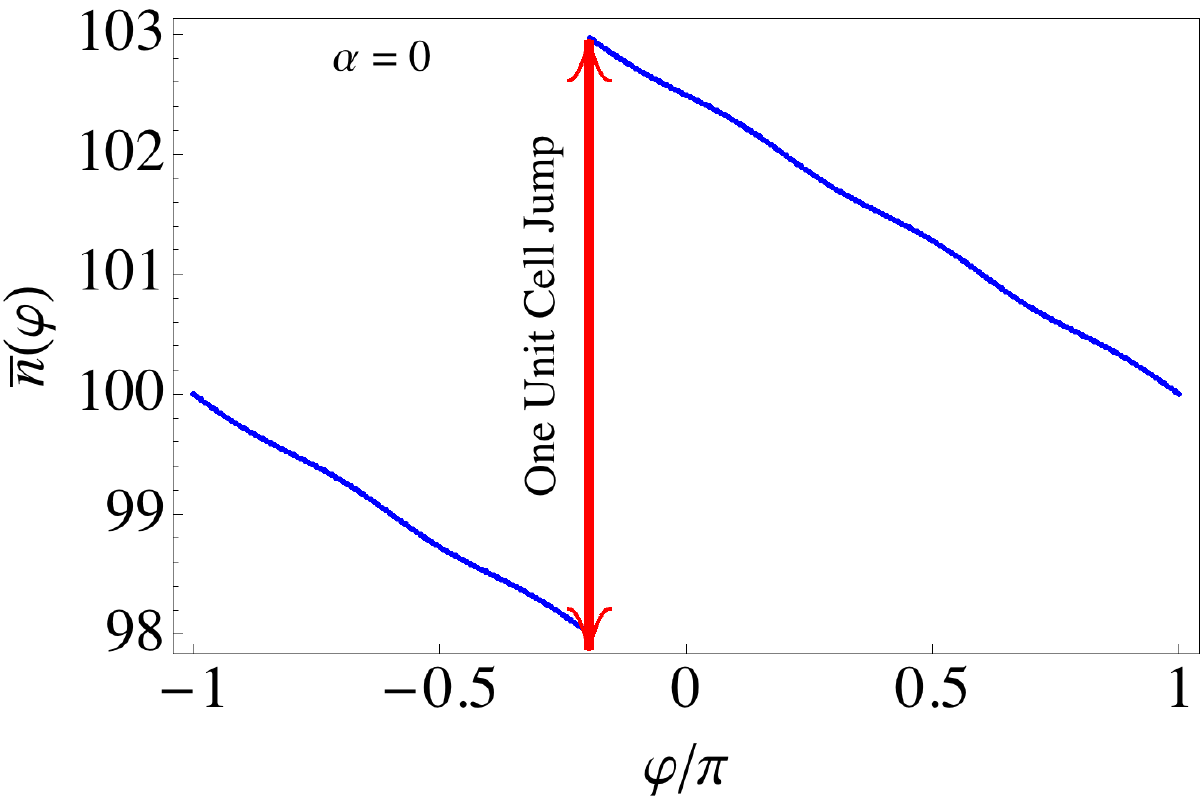}\hspace{0.1cm}\includegraphics[width=4.0cm,height=3cm]{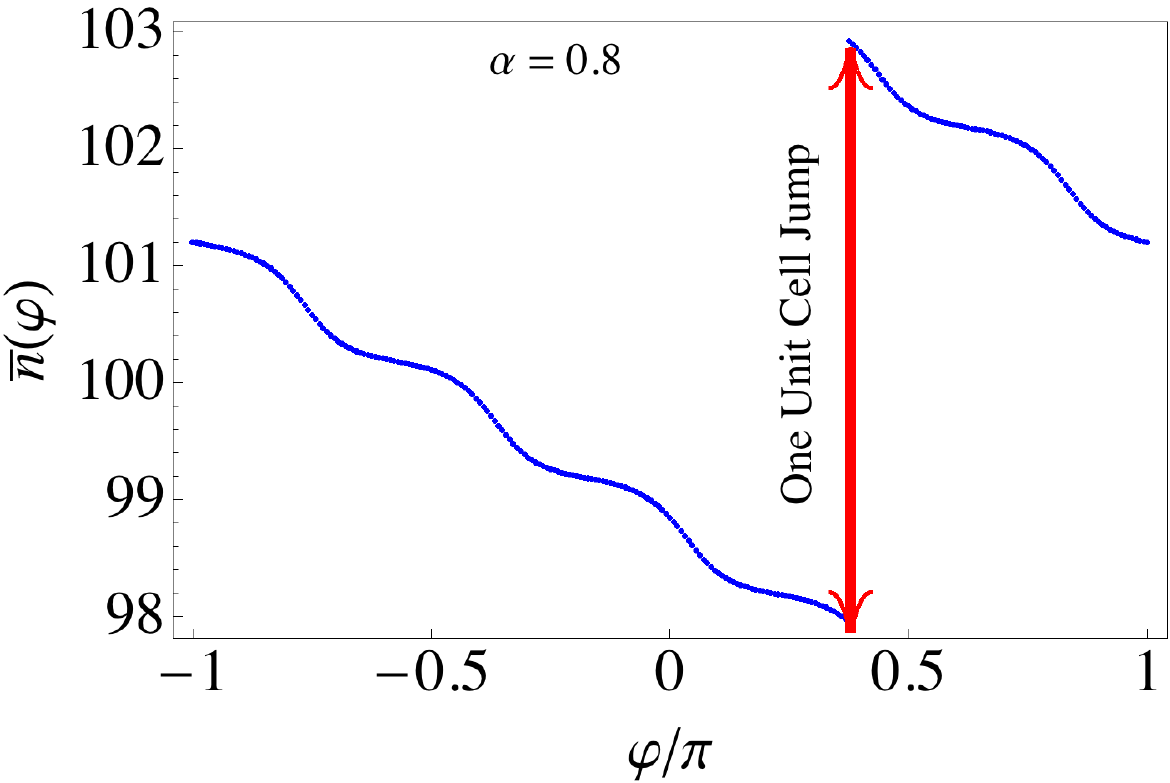}\\
\hspace{-0.2cm}
\includegraphics[width=4.0cm,height=3cm]{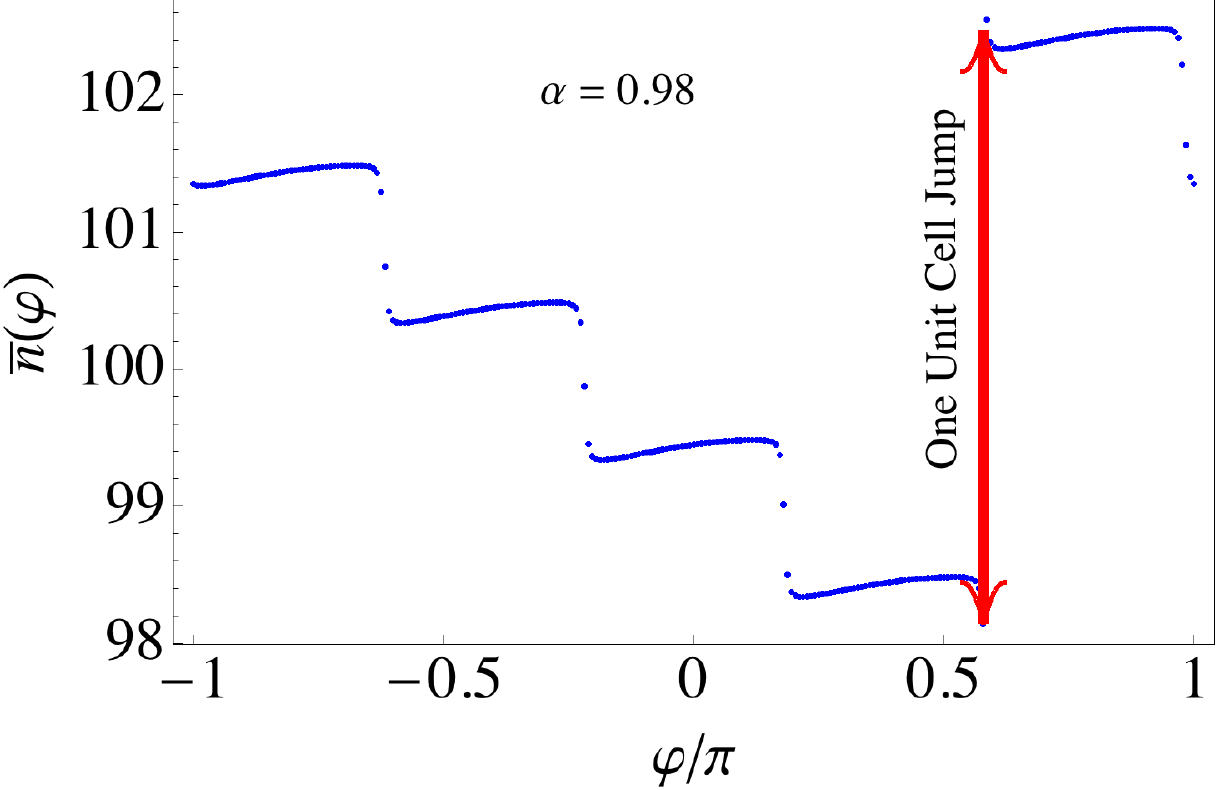}\hspace{0.1cm}\includegraphics[width=4.0cm,height=3cm]{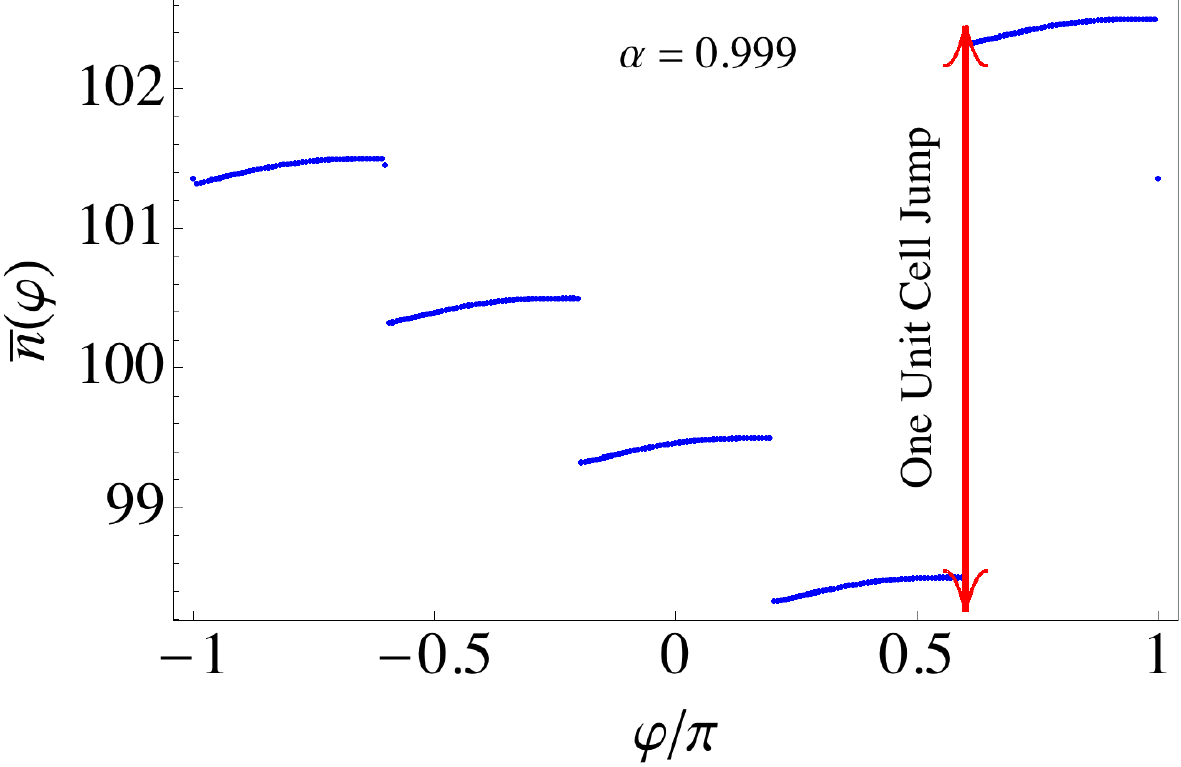}
    \caption{HWF centers plotted as a function of the adiabatic parameter $\varphi$ for $\alpha=0$ (AAH model),  $\alpha=0.8$,  $\alpha=0.98$ and  $\alpha=1.0$ (Maryland model).}
  \la{polarization}
\end{figure}

\textit{Chern number from polarization theory:-} 
In the following, we change $\alpha$ from 0 to 1 and track the change in the IQH topological invariant associated with the 2D system (Eq.~(\ref{eq:Hamiltonian})) in the hybrid space $(n,\varphi)$. An ideal tool for this task is the polarization of the 1D chain defined in the hybrid space~\cite{vanderbilt,restaprb}. 
The polarization of a finite 1D insulator is given by the average charge center of the hybrid Wannier function (HWF) ($\bar{n}(\varphi)$) of the system~\cite{troyer},
\begin{gather}
\bar{n}(\varphi)=\frac{\sum_{n} \langle n \rho(n,\varphi)\rangle}{\sum_{n}\langle\rho(n,\varphi)\rangle},\\
\rho(n,\varphi)=\sum_{\text{occupied states}} |n,\varphi\rangle\langle n,\varphi|,\nonumber
\end{gather}
where $n$ is the real space site index and $|n,\varphi\rangle$ is the hybrid eigenstate of the system, and the angular brackets $\langle...\rangle$ stand for the ground state expectation value given a fixed filling factor.

Non-zero Chern number is reflected in a discontinuity of $\bar{n}(\varphi)$ as a function of the phase (or gauge) parameter $\varphi$.
This discontinuity is a robust feature of the IQH 
and was recently proposed~\cite{troyer} as a tool to measure topological invariants directly in 2D cold atomic systems~\cite{ketterle,blochprl,mullervp}.
 	Note that the generalized 1D chain (Eq.~(\ref{eq:Hamiltonian})) has well defined gaps in the spectrum for any fixed $\varphi$ and $|\alpha|\leq1$ (including the Maryland model) which allows us to define the 1D polarization in terms of HWF centers in the whole parameter space. 
	
In Fig~\ref{polarization}, we plot the shift in the HWF centers for the same values of $\alpha$ (for $b=1/5$) as in Fig.~\ref{energy}.  We fix the filling factor (particle number per site) such that the chemical potential is in the gap above the top of the lowest band in the AAH limit ($\alpha=0$). 
In the limit of AAH, the HWF center as a function of $\varphi$ shows a one unit cell jump corresponding to the Chern number $C=1$, or, equivalently, a transfer of charge $e$ by a distance of one unit cell as $\varphi$ changes by a period, reflecting topological charge pumping~\cite{laughlin}. We monitor this jump (invariant) as we deform AAH ($\alpha=0$) to MM ($\alpha\rightarrow1$) keeping the filling factor fixed. Note that the polarization jump corresponding to the topological charge transfer survives in the MM limit $\alpha\rightarrow1$, see Fig.~\ref{polarization} bottom right. It may seem paradoxical at first that we can associate a Chern number with a gapless system. The limiting procedure $\alpha \rightarrow 1$ allows to project on to the states that are connected to the a topological band defined for $|\alpha|<1$. Note that the topology is not robust as any infinitesimal perturbation may mix the states thereby violating the quantization of the topological response. Such behavior is expected of a critical phase at $\alpha=1$ on general grounds. 
%
Note the additional discontinuities appearing in HWF shift $\bar{n}(\varphi)$ in the case of MM, Fig.~\ref{polarization} bottom right arise due to the divergent onsite potential effectively breaking the system up into smaller subsystems coupled by tunneling. 

 \begin{figure}[htb!]
  \centering
\includegraphics[width=4cm,height=3cm]{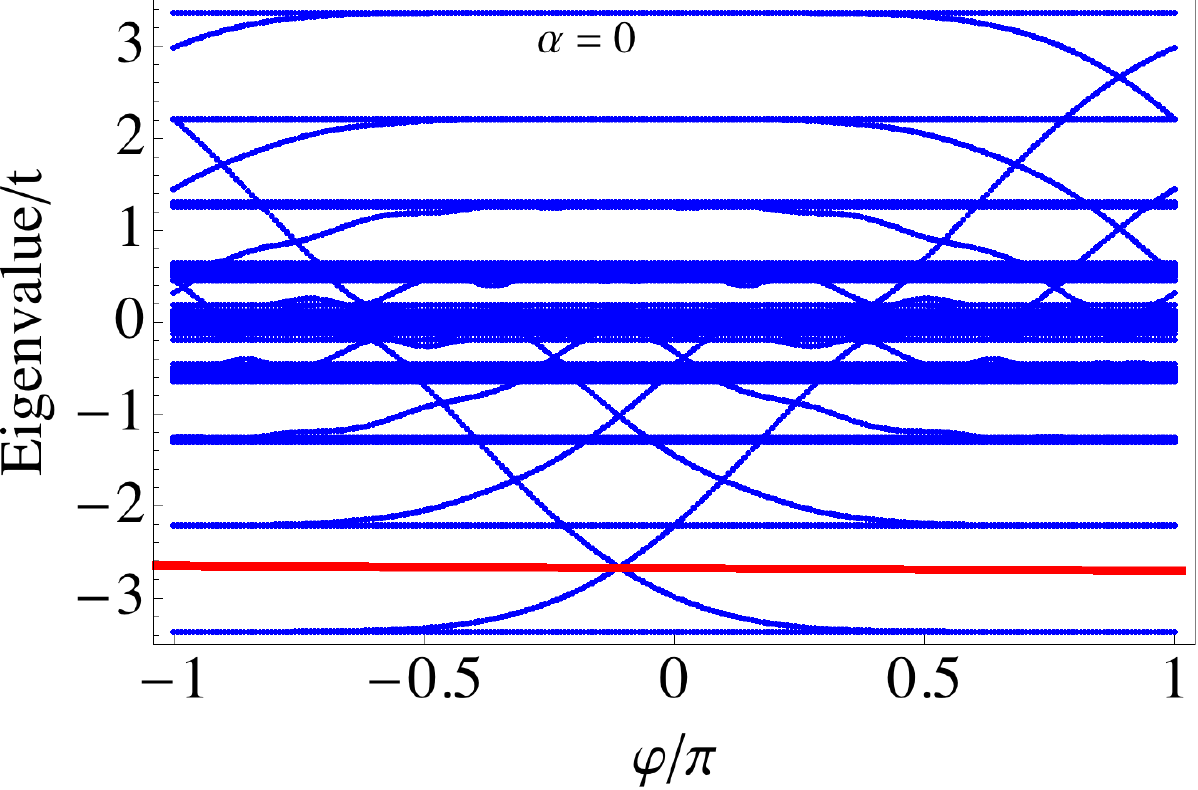}\hspace{0.1cm}\includegraphics[width=4.0cm,height=3cm]{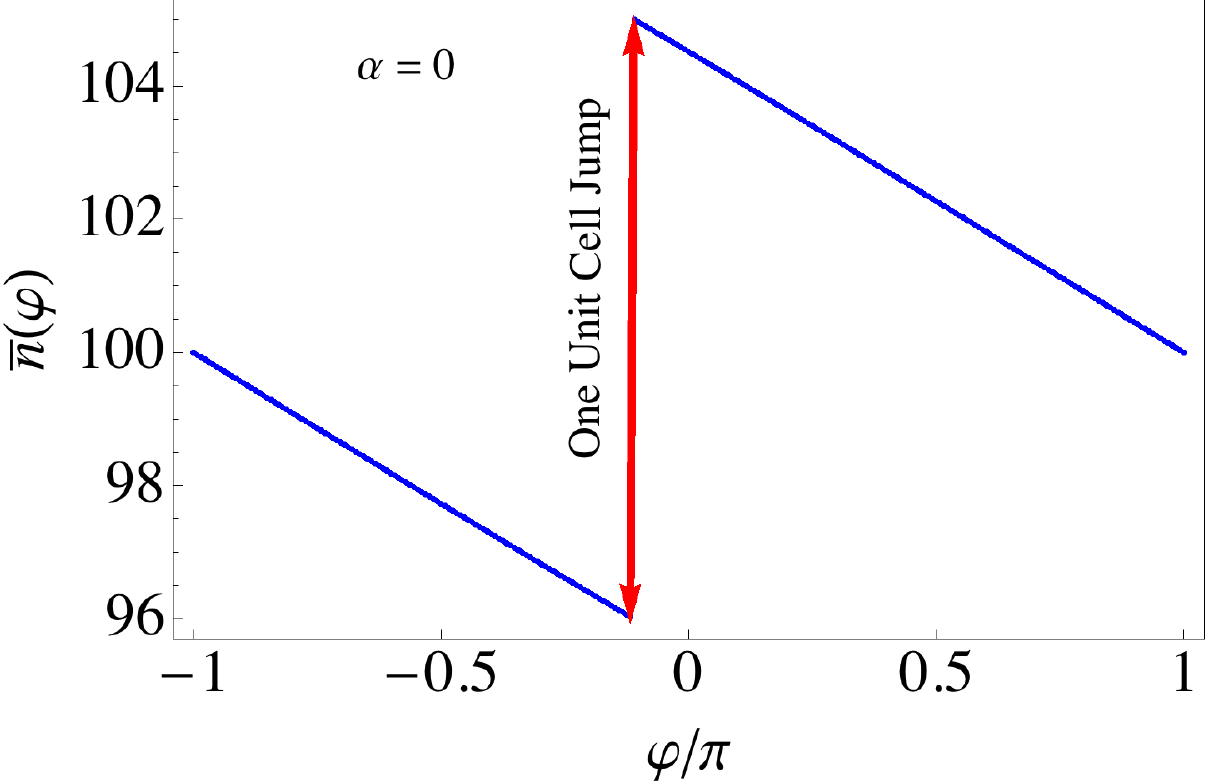}\\
\hspace{-0.2cm}
\includegraphics[width=4.0cm,height=3cm]{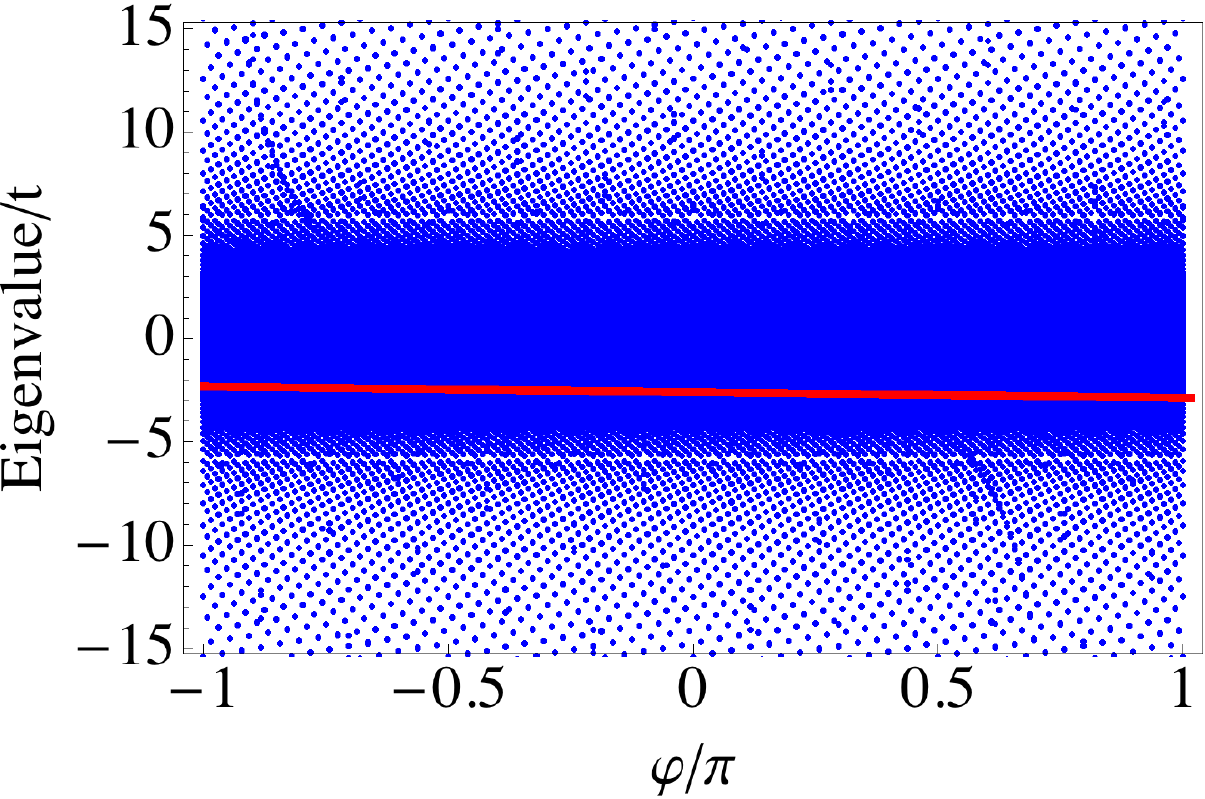}\hspace{0.1cm}\includegraphics[width=4.0cm,height=3cm]{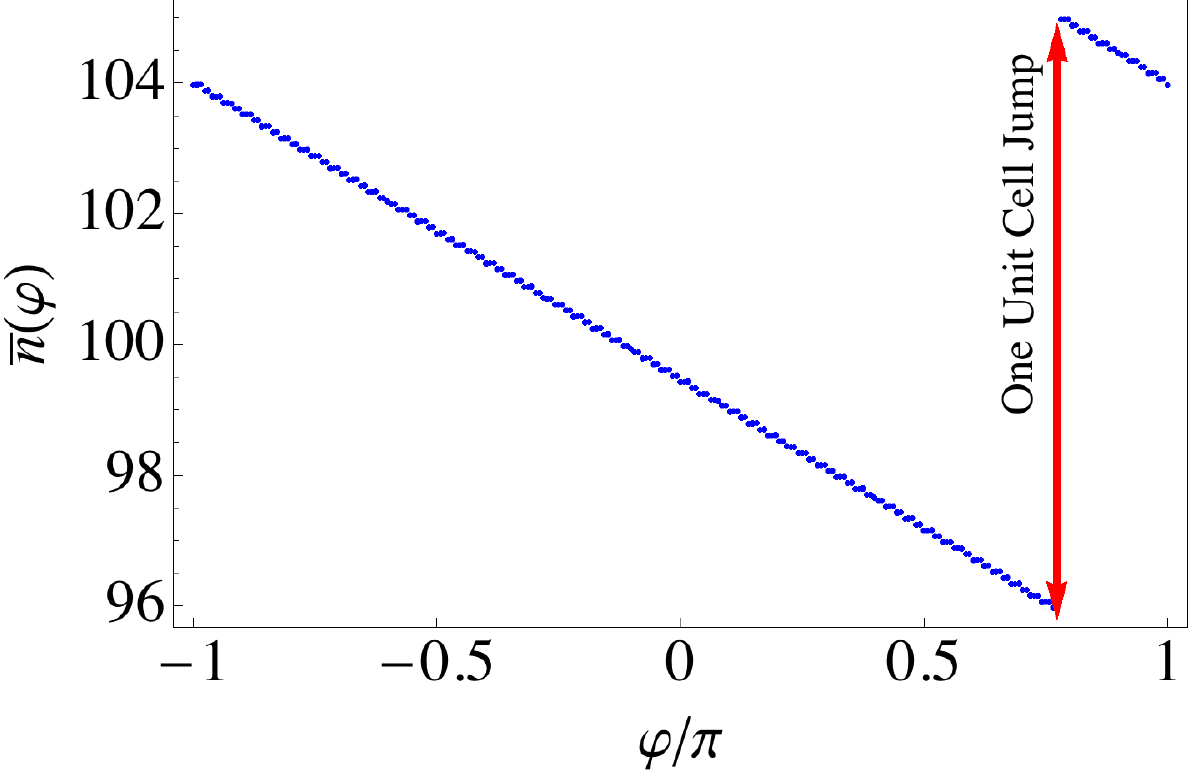}    \caption{1D quasicrystal band structure and the shift in polarization as a function of the phase $\varphi$ for AAH model (upper panel) and Maryland model (lower panel) for $b=\frac{110001}{1000000}$.   }
  \la{quasicrystal}
\end{figure}

\textit{Topological classification of 1D quasicrystals:-}
Families of 1D incommensurate tight binding models manifest a special `quasiperiodic' translational invariance: 
an arbitrary shift in the phase $\varphi\rightarrow\varphi+\delta\varphi$ can be compensated by a shift along the chain $n\rightarrow n+\delta n_{\delta\varphi}$. Note that this is true only at irrational values of $b$ since only in this case $2\pi b n$ forms a dense set mod $2\pi$. 
It has been argued~\cite{kraus1} that this quasiperiodic translational invariance allows one to assign the same Chern number to each member of the family of QCs, i.e. for each value of the phase parameter $\varphi$.  This interpretation has been challenged in Ref.~\cite{brouwer}.
The quasiperiodic translation symmetry is preserved in the case of the Maryland model ($\alpha=1$) which sits exactly at the critical point of a 2D TQPT. 
\begin{figure}[htb!]
  \centering
  {\includegraphics[scale=0.4]{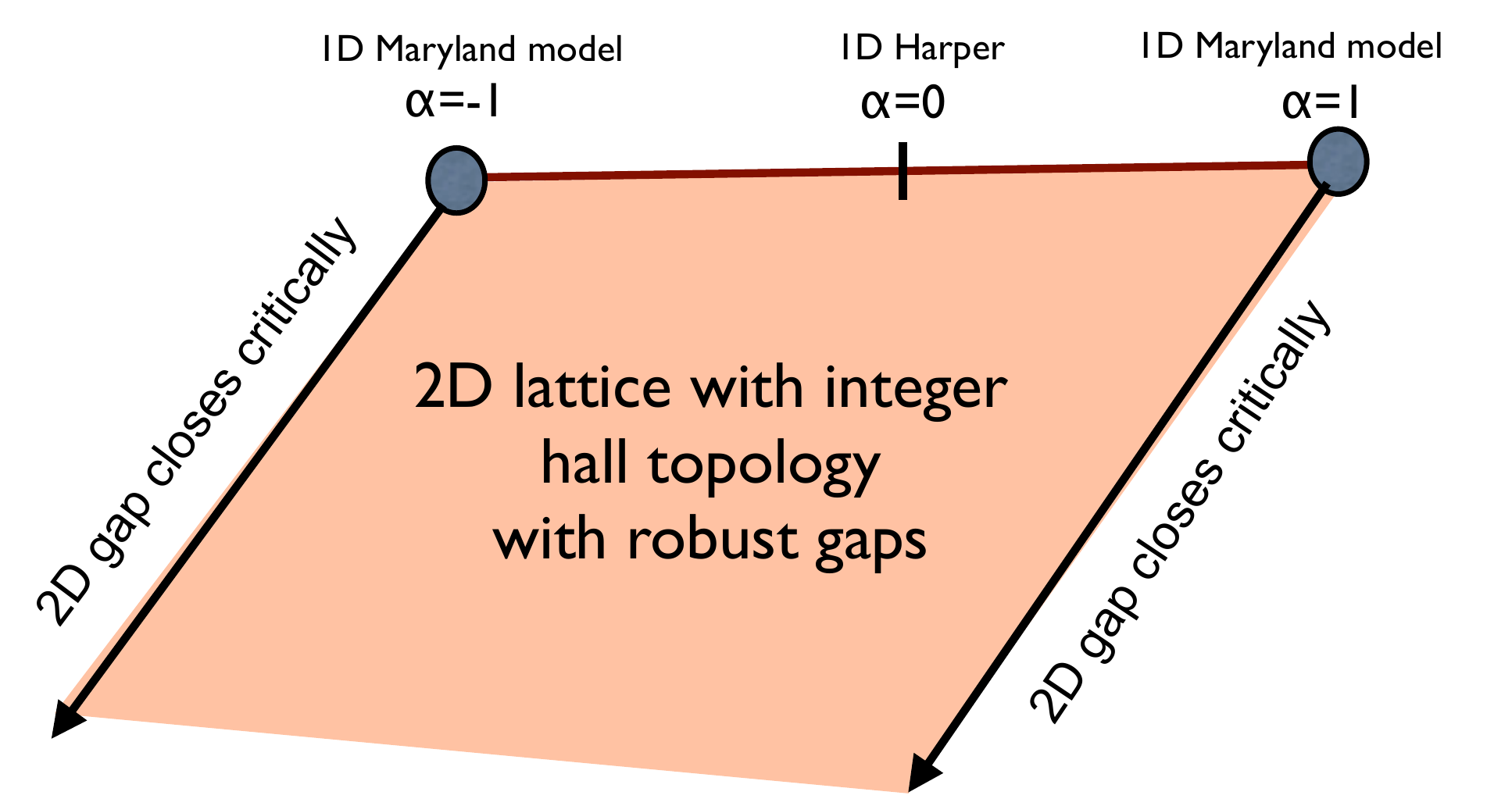}}
  \caption{The phase diagram of Eq.~(\ref{eq:Hamiltonian}) parameterized by $|\alpha|\leq1$, the deformation parameter interpolating between the AAH and the Maryland model.}
  \label{phase}
\end{figure}
%
In Fig.~\ref{quasicrystal} we plot the band structure and the change in the polarization as a function of the phase $\varphi$ for the incommensurate AAH and MM. We choose the flux fraction to be a truncated Liouville constant (Liouville numbers are irrational numbers infinitely close to rational numbers). Note that the finite size Maryland model still demonstrates the presence of a non-zero Chern number in the same restricted sense as we found for the commensurate case. The constant slope of $\bar{n}(\varphi)$ in Fig.~(\ref{quasicrystal}) manifests the constant Berry curvature (as a function of $\varphi$), see~\cite{supp} for details. The latter being a signature of the `quasiperiodic' translation invariance as noted by Kraus et. al in Ref.~\cite{kraus1}. 
%
Remarkably, the spectrum is gapped in the incommensurate (Liouville) 1D model Eq.~(\ref{eq:Hamiltonian}) (for fixed $\varphi$) for $|\alpha|<1$ and forms a dense set for $\alpha=\pm1$ (rather than a continuous set),  whereas the equivalent 2D model becomes gapless as we approach critical points $\alpha=\pm1$. The details of the 1D spectrum depend on the type of the irrational number $b$ however at no value the spectrum becomes absolutely continuous~\cite{simon}.    
Within the class of 1D models with quasiperiodic symmetry, Maryland model manifests a 2D topological phase transition as a function of the deformation parameter $\alpha$ which can only be realized by sweeping the phase $\varphi$ (see Fig.~\ref{phase}). 

 \textit{Conclusion:-} We have identified a previously unknown topological feature of the Maryland model introduced in Refs.~\cite{grempel,prange} in the context of Anderson localization and kicked quantum rotor studies. We show that this model represents a topological quantum phase transition point in a class of corresponding 2D lattice models with IQH topology. The criticality allows us to associate topological invariants with the Maryland model in a restricted mathematical sense at the special filling factors that are adiabatically connected to the spectral gaps in the 1D Aubry-Andre-Harper model. Our theory presented in this work establishes deep and unexpected mathematical connections between 2D topological models and a family of 1D incommensurate localization models. 
 
\textit{Acknowledgements:-}We would like to thank Michael E. Fisher, Shmuel Fishman, Zohar Ringel and A. G. Abanov for stimulating discussions. We also thank Edwin Barnes for careful reading of the manuscript. This work is supported by JQI-NSF-PFC, Microsoft Q and JQI-ARO-MURI.




\appendix
\section{Appendix}
\subsection{2D ancestor for general model}
In this section we give details of obtaining the 2D real space lattice starting from 1D Hamiltonian (Eq.~(\ref{eq:Hamiltonian})). We start by inverse fourier transforming the momentum parameter of the second dimension $\varphi$. This is a standard procedure that has been carried out in a compact way in Ref.~(\cite{kraus2}). This exercise allows us to capture the 2D lattice with flux as a function of the deformation parameter $\alpha$. 
\begin{gather}
H(\varphi,\alpha)=-\sum_{n=1}^{N-1} t (c^{\dagger}_{n+1}c_n+c^{\dagger}_n c_{n+1})-\sum_{n=1}^{N} V_n(\alpha,\varphi)c^{\dagger}_{n}c_n,\la{eq:Hamiltonian}\\
V_{n}(\alpha,\varphi)=2\text{\ensuremath{\lambda}}\left(\frac{\cos\left(2\pi n b+\varphi-\alpha\frac{\pi}{2}\right)}{1+\alpha\ \cos(2\pi n b+\varphi)}\right),\nonumber
\end{gather}
The real space 2D lattice equivalent of the above 1D Hamiltonian can be obtained from the following fourier transform.
\bea
H_{2D}&=&\int_{-\pi}^{\pi}d\varphi H(\varphi)e^{in\varphi}\nonumber\\
c_{n}\equiv c_{n}(\varphi)&=&\frac{1}{\sqrt{2\pi}}\sum_{m}e^{i\varphi m}c_{n,m}
\eea
We can explicitly write out the 2D real space lattice as,
\bea
H_{2D}&=&\sum_{m,n}[-t(c_{n,m}^{\dagger}c_{n+1,m}+c_{n+1,m}^{\dagger}c_{n.m})\\
&+&\lambda\sum_{m'}\int\frac{d\varphi}{2\pi}\frac{-V_{n}(\alpha,\varphi)}{2}(e^{i\varphi(m-m')}c_{n,m}^{\dagger}c_{n,m'}+h.c.)]\nonumber
\eea
we can perform the shift $m'=m-l$ where $l$ can be defined as the range of hopping 
\bea
H_{2D}&=&\sum_{m,n}[-t(c_{n,m}^{\dagger}c_{n+1,m}+c_{n+1,m}^{\dagger}c_{n.m})\\
&-&\lambda\sum_{l}I_{nl}(e^{i\varphi l}c_{n,m}^{\dagger}c_{n,m-l}+h.c.)]\nonumber
\eea
Now all that remains is to calculate the hopping terms $I_{nl}$ in the second term which is given in the form of an integral.
\[
I_{nl}=\int\frac{d\varphi}{2\pi}\left(\frac{\cos(2\pi nb+\varphi-\alpha\frac{\pi}{2})}{1+\alpha\cos(2\pi nb+\varphi)}\right)e^{i\varphi l}
\]
Following Ref.~(\cite{kraus2}), we perform analytical continuation to the complex plane by setting $z=e^{i\varphi}$. The 2D hopping term can be written as an integral over an unit circle in the complex plane.
\[
I_{nl}=e^{-il(2\pi nb+\alpha\frac{\pi}{2})}\oint\frac{dz}{2\pi i}\ z^{l-1}\frac{z^{2}+e^{i\alpha\pi}}{\alpha z^{2}+2z+\alpha}
\]
For the above contour integral there are simple poles for any $l\ne0$ at $z=\frac{-1\pm\sqrt{1-\alpha^{2}}}{\alpha}$ and for $l=0$ (onsite energy term) there is an additional pole at $z=0$. Since the contour is defined on a unit circle, the poles within the unit circle are $z=0\ (l=0)$ and $z=\frac{-1+\sqrt{1-\alpha^{2}}}{\alpha}$.
\bea
I_{nl}(\alpha)&=&e^{-il(2\pi nb+\alpha\frac{\pi}{2})}\bigg[\frac{e^{i \pi  \alpha }}{\alpha }\delta _{l,0}+\left(-1+\sqrt{1-\alpha ^2}\right)^{l-1}\nonumber\\
&\times&\frac{ \left(2-\alpha ^2(1-e^{i \pi  \alpha } )-2 \sqrt{1-\alpha ^2}\right)}{2 \alpha ^{l+1} \sqrt{1-\alpha ^2} }\bigg].
\la{hop}
\eea
At the special points of the parameter $\alpha\rightarrow0$ (AAH limit), we have $I_{n1} =  1$ and for $l\ne0$ $I_{nl}  =  0$. In this limit we recover the 2D lattice equivalent of AAH model, which is the nearest neighbor 2D lattice with a flux. For $\alpha\rightarrow1$ (MM limit) the poles approach the unit circle contour and has to be evaluated via principal value prescription. In this limit we obtain an infinite range hopping in the m direction. This non-local hopping manifests in the hopping term as  $I_{n0}= 0$ and $I_{nl} = (-1)^{l}\forall l\ge1$.

\subsection{Universal scaling of the energy gaps near the critical point $\alpha\rightarrow1$}
\begin{figure}[htb!]
  \centering
\includegraphics[width=8cm,height=5cm]{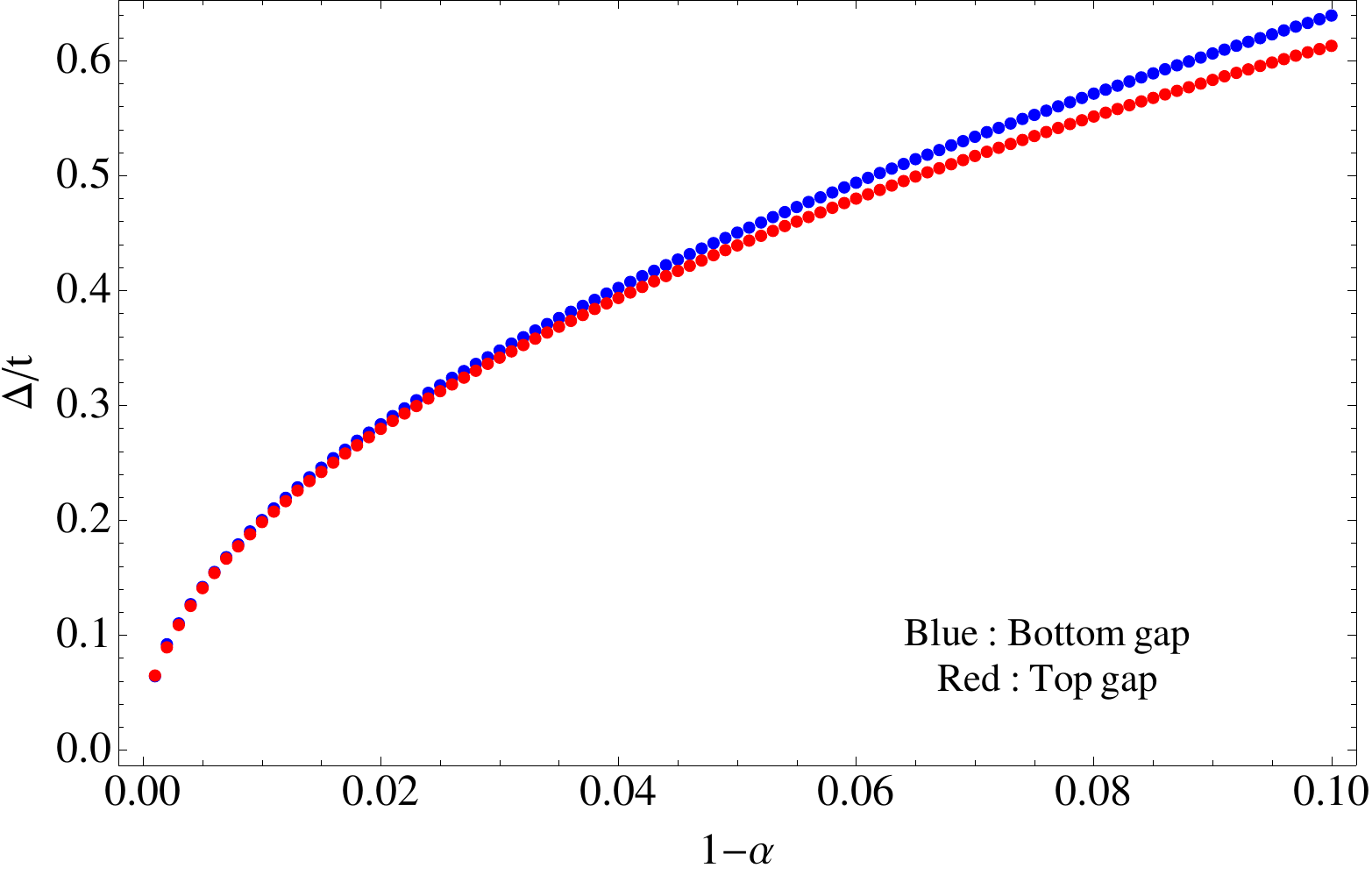}
  \caption{Scaling of the Band gaps for b=1/3 as a function of $1-\alpha$. Red and Blue lines correspond to the two gaps for the case of b=1/3. Numerical energy spectrum is calculated for periodic boundary conditions and the number of sites $N=99$.}
  \label{gapclosing}
\end{figure}
In this section, we provide additional details to shed further light on the nature of the critical phenomena in the generalized model defined in Eq.~(1). It is instructive to monitor the behavior of a physical characteristic of the system as the critical point is approached in the parameter space. In this case the physical quantity we look at is the energy gap, which is inversely related to some power of the correlation length in the system. The gap vanishes in the limit $\alpha \rightarrow1$ of the critical point. Monitoring the band gap as a function of $\alpha$ has a twofold advantage.: 1) We demonstrate the universal scaling of the energy gap close to the critical point by monitoring different gaps.  We show that the gaps converge to the same critical behavior as we get approach the critical point. This is provides additional evidence for the underlying scale invariance and associated criticality of the system at $\alpha=1$. 2) We can extract critical exponents that may be used to identify the universal scaling behavior of the critical phenomena. For the general 2D model described in the main text in the hybrid representation $(n,\varphi)$, the energy gaps are indirect. Each of the gaps gets progressively smaller as $\alpha \rightarrow 1$ and closes at exactly $q$ points in the full 2D Brillouin zone (as opposed to the reduced Brillouin zone defined by magnetic translations), given a rational flux fraction $b=p/q$. We monitor the energy gap as a function of the deviation from criticallity characterized by $1-\alpha$ which we define as 
\begin{equation}
\Delta =(Min \{E_{n+1}\}-Max\{E_{n}\}),  
\end{equation}
where $E_n$ is the energy of the $n$th subband in the spectrum. For $b=p/q$, there are $q-1$ gaps separating $q$ subbands in the spectrum~\cite{Hofstadter}. Note that the two terms in the right hand side may be at different different points in the 2D Brillouin zone reflecting the indirect nature of the gap.
\begin{figure}[htb!]
  \centering
\includegraphics[width=8cm,height=5cm]{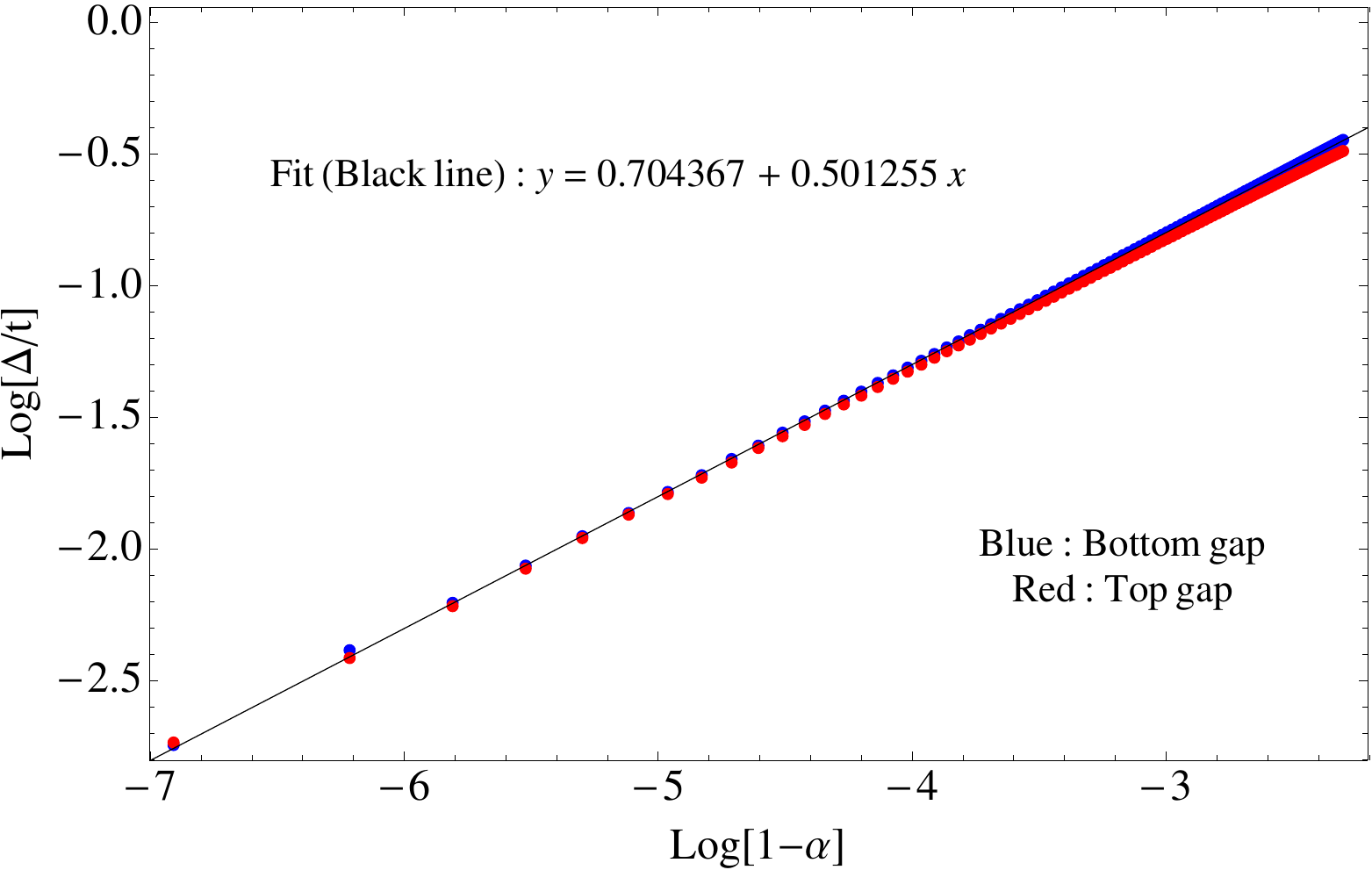}
  \caption{Log-Log plot of the gap vs $1-\alpha$. Red and Blue points correspond to numerical data and black line is a linear fit of the numerical data. Numerical energy spectrum is calculated for periodic boundary conditions and the number of sites $N=99$.}
  \label{exponent}
\end{figure}
We consider an example where $b=1/3$. The spectrum has two gaps in the AAH limit $\alpha\rightarrow0$. In the MM limit the bands touch at 3 points where the gaps close. We impose periodic boundary conditions for $N=99$ sites. In Fig.~(\ref{gapclosing}) we plot the energy difference $\Delta$ between the extreme points in the subbands at these gap closing points. We plot $\Delta(n)$ for the two gaps ($n=1$ and $n=2$) as a function of $1-\alpha$ for $\alpha$ between 0.900-0.999 in steps of 0.001. We extract the critical exponent from the slope of the log-log plot (see Fig.~(\ref{exponent})) by fitting to a straight line.  The numerical critical exponent we obtain is $\sim0.5$, which indicates a mean field type of critical behavior.
\subsection{Exact energy spectrum for Maryland Model for $b=1/3$.}
\begin{center}
\begin{figure}[htb!]
\textrm{(a)}\hspace{4cm}\textrm{(b)}
\includegraphics[width=4cm,height=3cm]{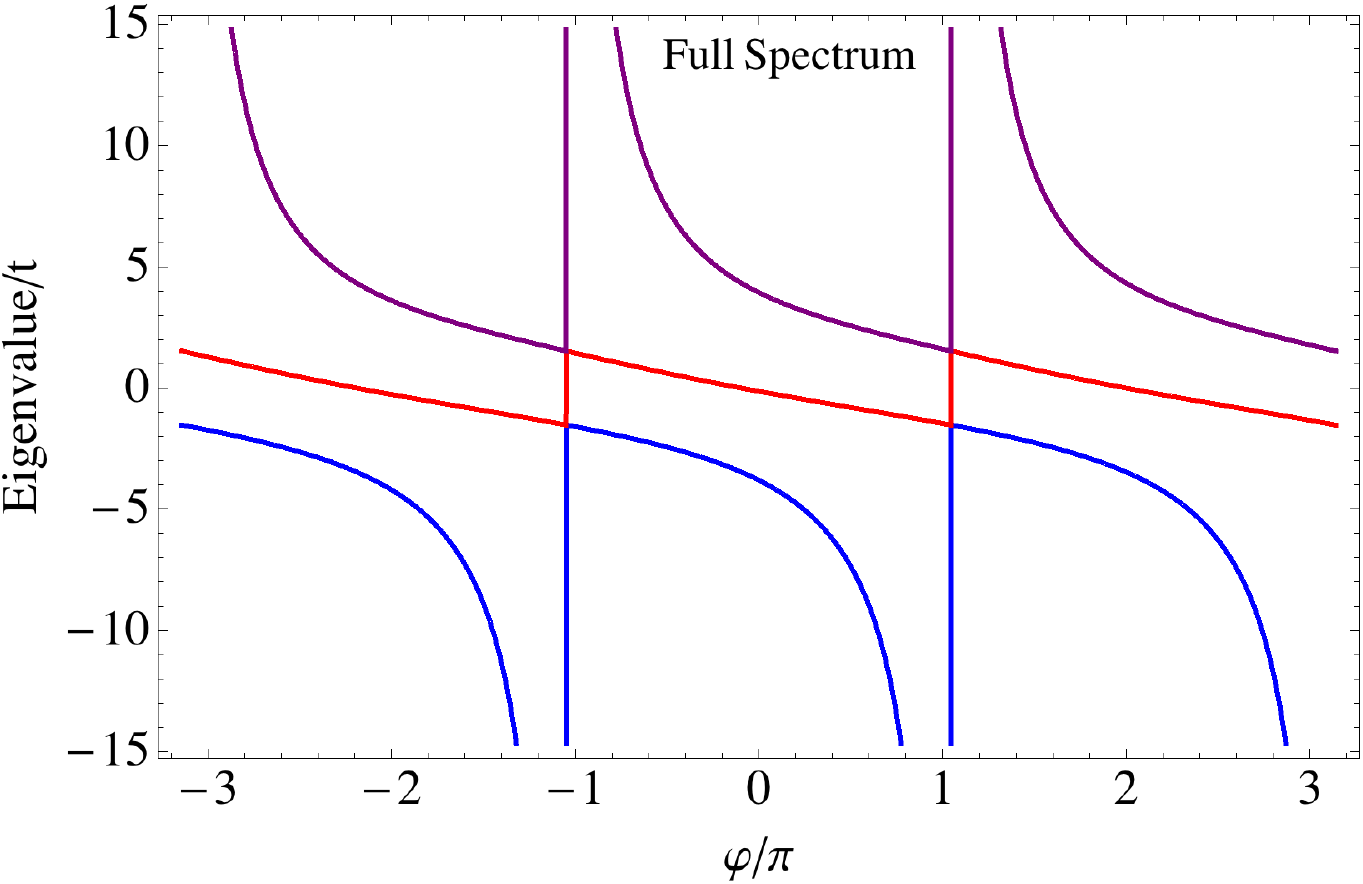}\hspace{0.1cm}\includegraphics[width=4.0cm,height=3cm]{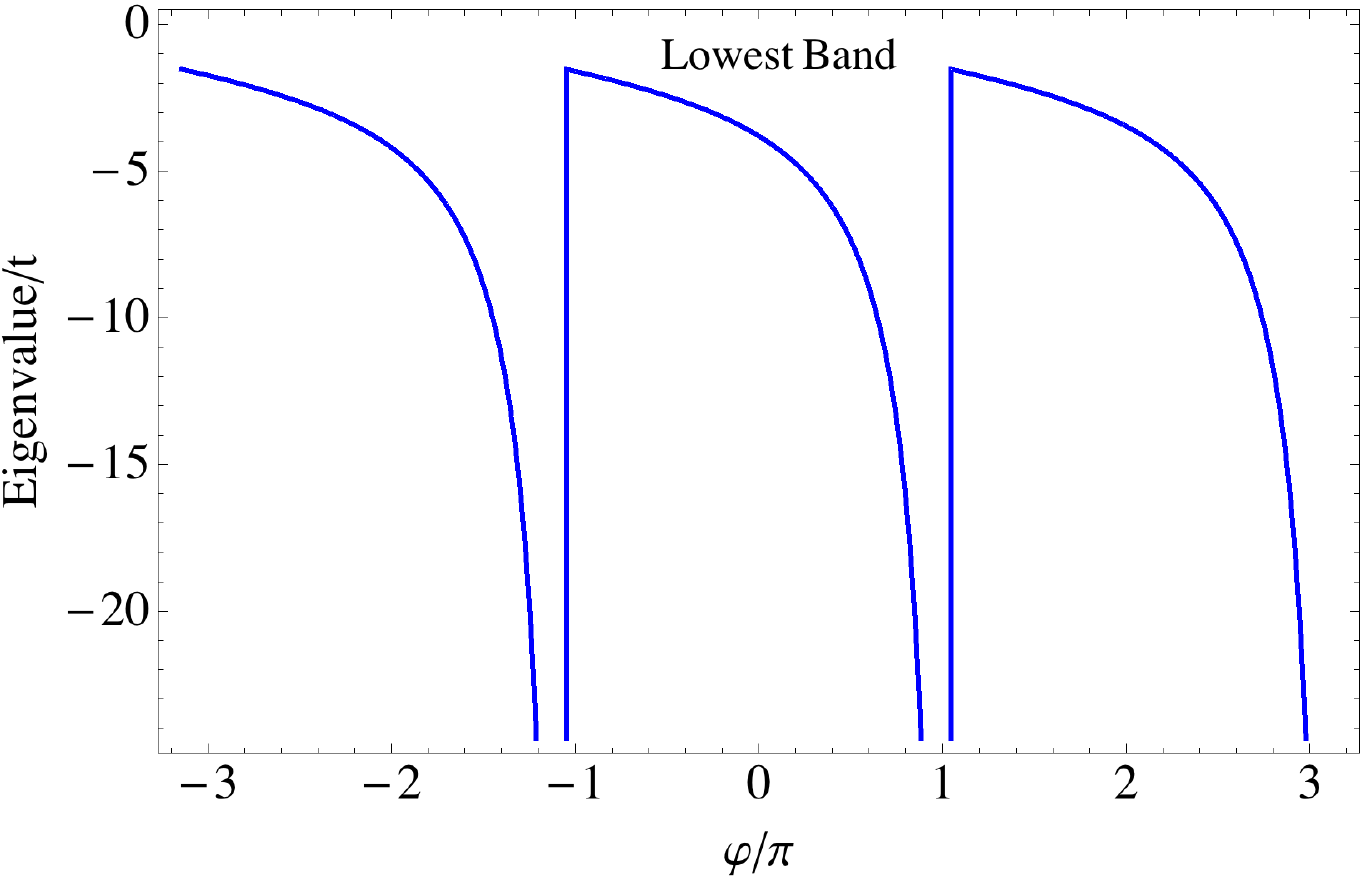}\\
\hspace{-0.2cm}\textrm{(c)}\hspace{4cm}\textrm{(d)}
\includegraphics[width=4.0cm,height=3cm]{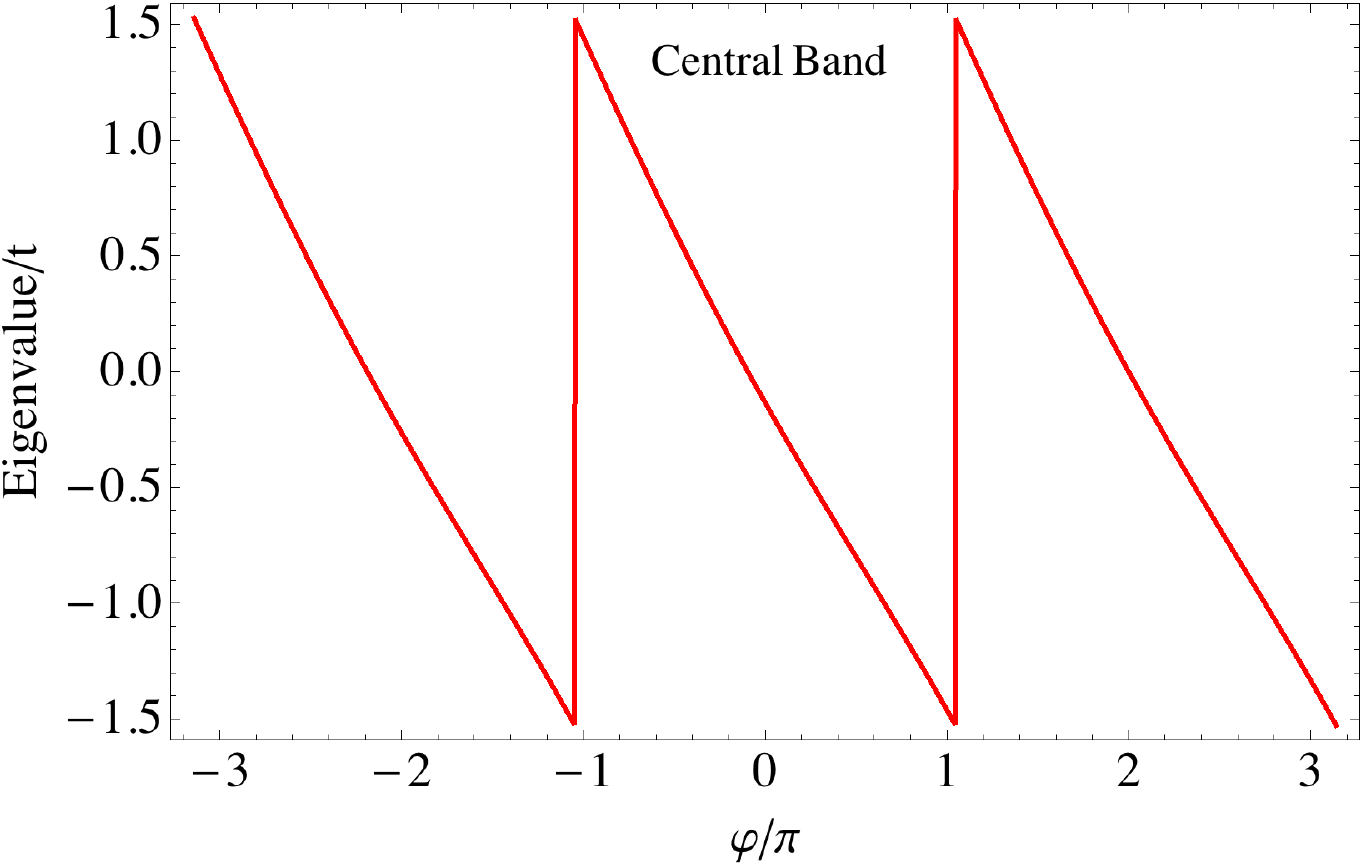}\hspace{0.1cm}\includegraphics[width=4.0cm,height=3cm]{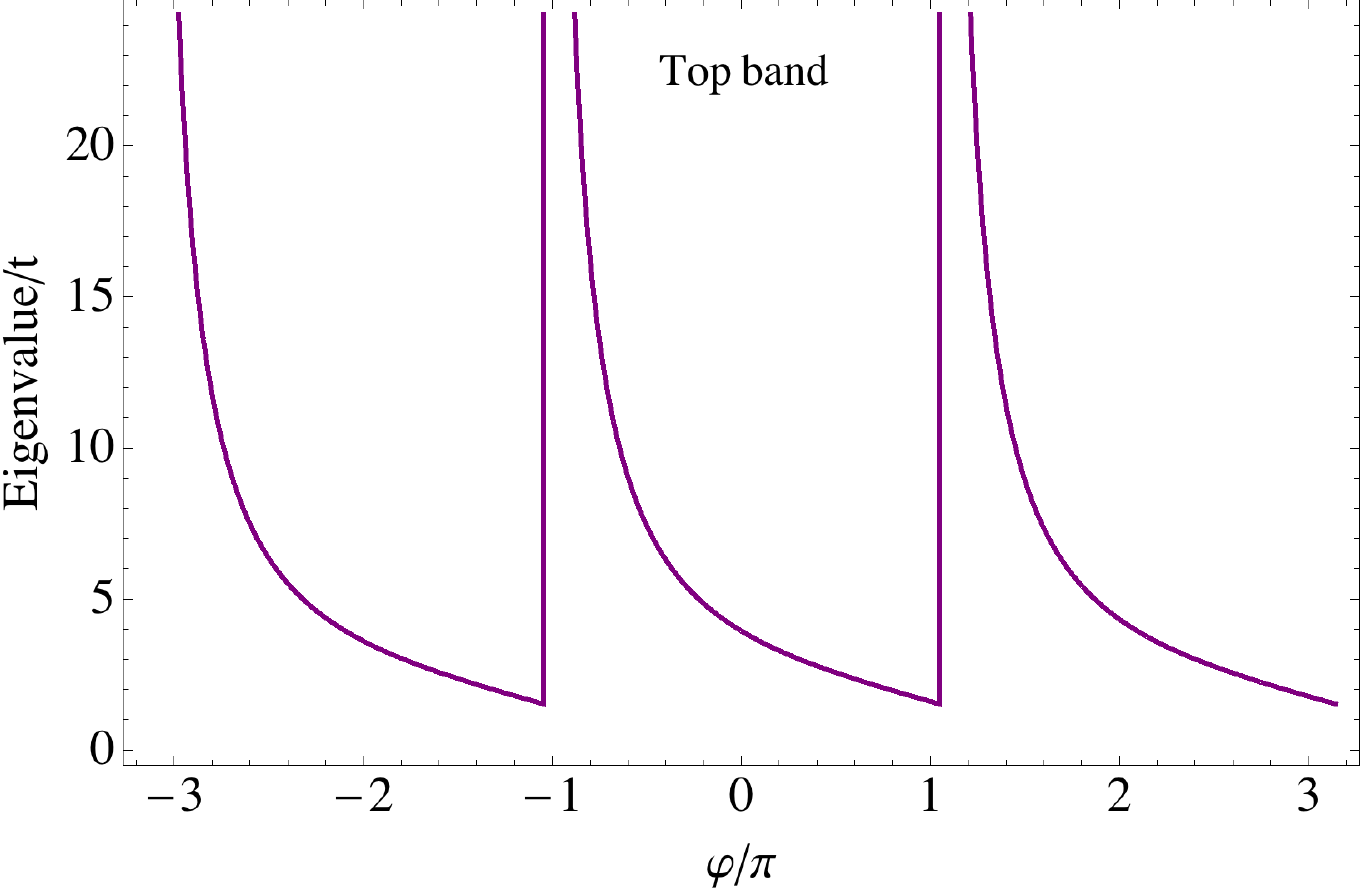}
\caption{Exact energy spectrum with periodic boundary conditions.(a) Energy spectrum as a function of the parameter $\varphi$ with fixed $\theta=0,$ $\lambda=1$ for $b=1/3$. (b) Spectrum of the solution corresponding the lowest energy subband. (c) Energy spectrum corresponding to the central subband. (d) Exact energy Spectrum corresponding to the top band. }
\label{energy}
\end{figure}
\end{center}
In this section we present the exact energy spectrum of the commensurate MM for flux fraction $b=1/3$. Following Ref.~(\cite{watson}), we can write a general equation for the energy spectrum for arbitrary values of $b=p/q$. 
\begin{gather}
\cos qk_x=\cosh q\gamma \cos q\mu+\sinh q\gamma\sin q\mu \cot(q(\varphi+\pi/2))\la{spectrum}
\end{gather}
where we have defined $\cosh (\gamma) \cos(\mu)=E/2$  and $\sinh(\gamma)\sin(\mu)=\lambda$. E is the energy eigenvalue and $k_x$ is the fourier image of the lattice site index $n$. Using Eq.~(\ref{spectrum}), one can analytically compute the energy spectrum $E\equiv E(k_x,\varphi)$. Eq.~(\ref{spectrum}) immediately shows that there are $q$ solutions corresponding to $q$ subbands. For $b=1/3$, the solutions can be obtained in closed form. The actual expression are cumbersome and we just show plots of the spectrum solutions for $b=1/3$ (Fig.~(\ref{energy})) for $k_x=0$, which corresponds to the smallest gap along $k_x$. We plot three solutions corresponding to the three bands Fig.~\ref{energy}\ b,c,d. We put together these bands to construct the full spectrum Fig.~\ref{energy}a. The full spectrum is in agreement with the numerical results in the main text in the MM limit. Note however that the numerical calculation was done for open boundary conditions and therefore manifests edge states. The exact spectrum demonstrates the gapless nature of the 2D spectrum of MM. The energy gaps between subbands close exactly at $\alpha=1$ and as a result there is no mixing between the states of in different subbands except exactly at the point of subband touching (which is a set of measure zero). This allows us to define the topological index for each subband in a mathematical sense by projecting onto the states of the lowest band up to the touching point, i.e. the filling factor is such that the chemical potential is exactly at the touching point. Clearly any attempt to measure response in this case will excite states from both of the subbands and therefore the response function will not be quantized. In this sense topological index is not well defined.  
\subsection{Periodic vs open boundary conditions}
\begin{center}
\begin{figure}[htb!]
\textrm{(a)}\hspace{4cm}\textrm{(b)}
\includegraphics[width=4cm,height=3cm]{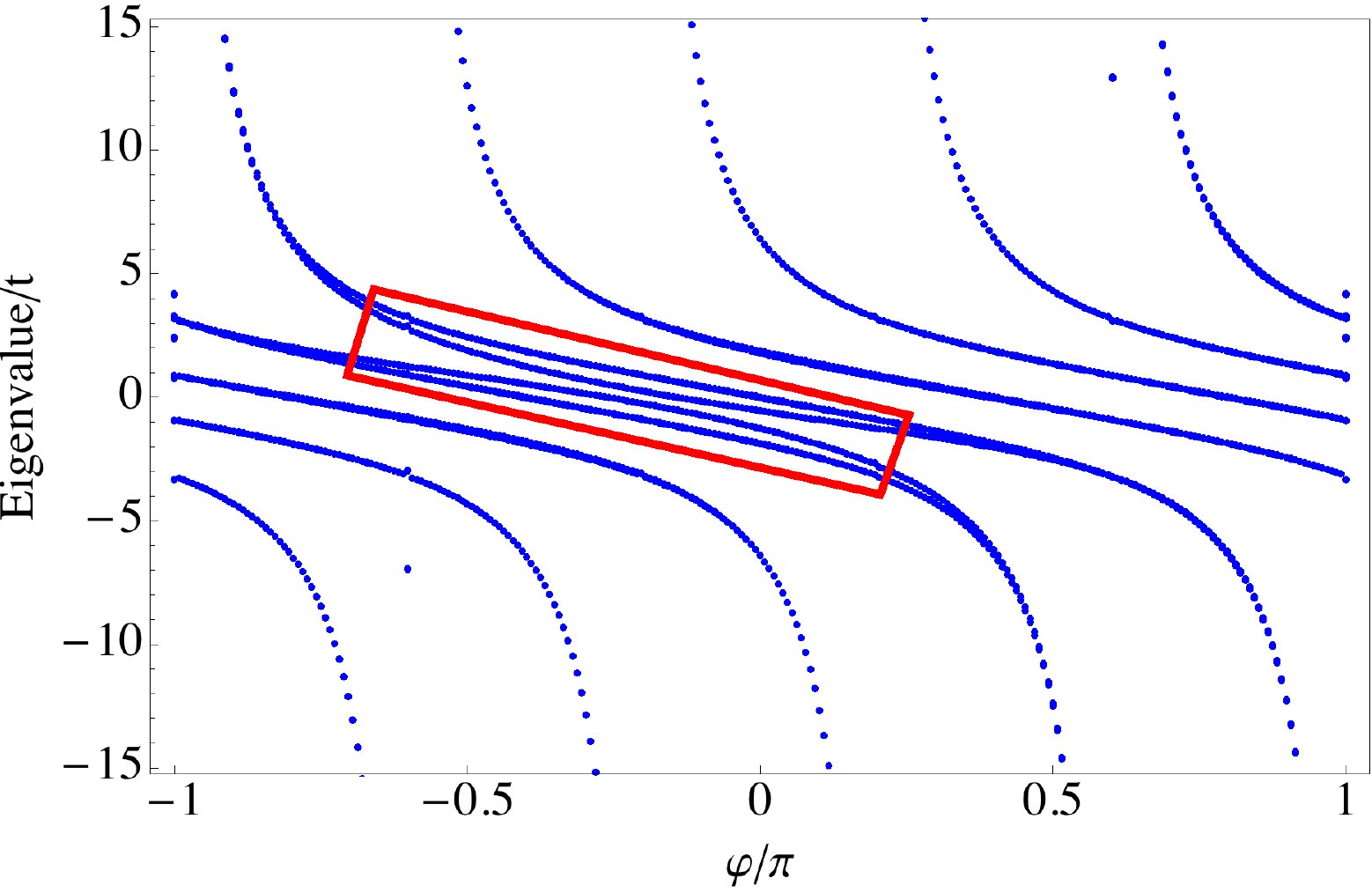}\hspace{0.1cm}\includegraphics[width=4.0cm,height=3cm]{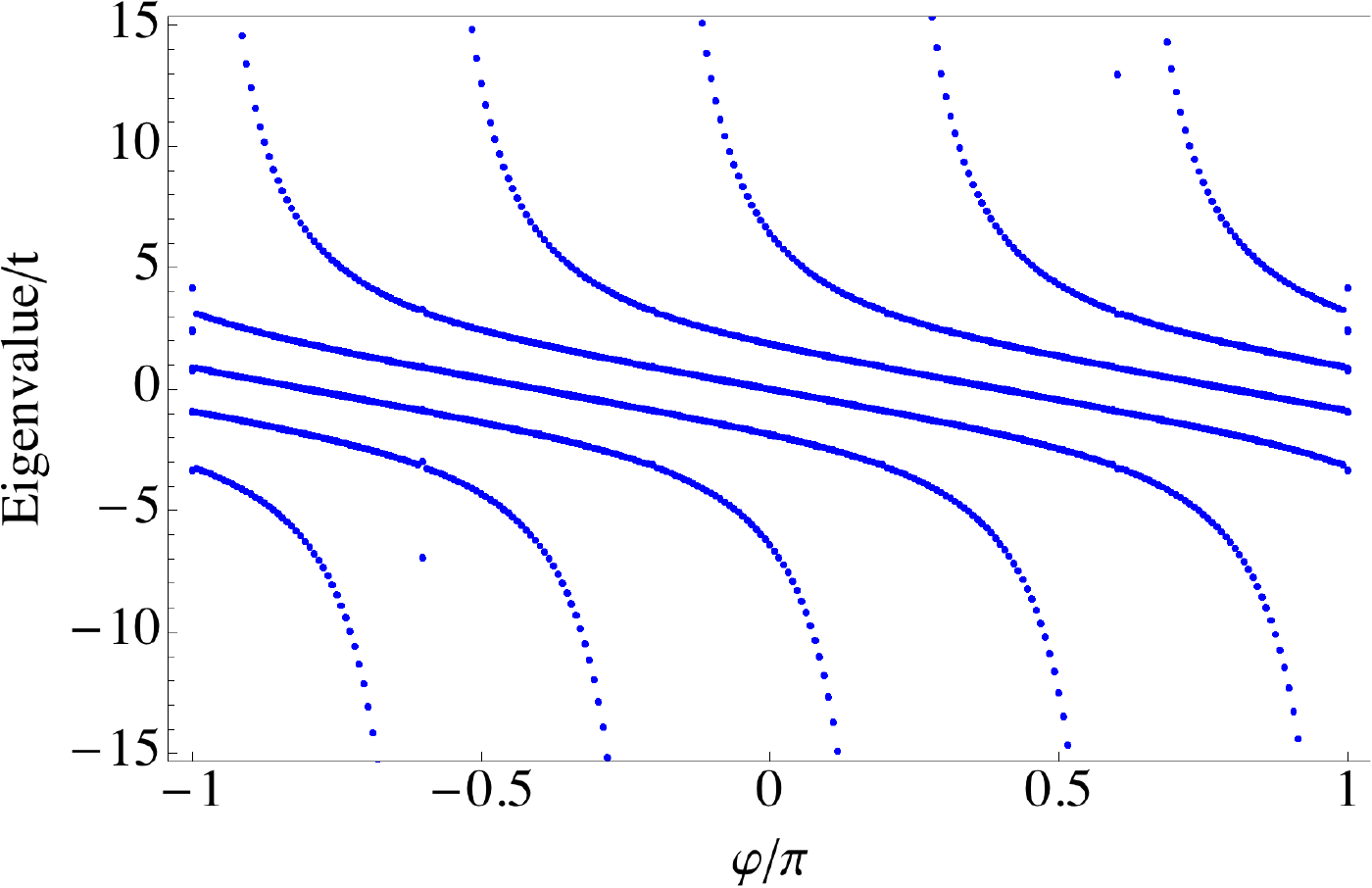}\\
\caption{Numerical energy spectrum as a function of $\varphi$ with (a) open boundary conditions and (b) periodic boundary conditions for $\alpha=0.999,$ $N=200$ sites and $b=1/5$. } \label{pbc}
\end{figure}
\end{center}

In this section we show the numerical energy spectrum for periodic and open boundary conditions in the MM limit ($\alpha=0.999$). Fig.~(\ref{pbc}a) shows the energy spectrum with open boundary conditions (obc) for $N=200$ sites and $b=1/5$. The edge states manifest for obc (marked within the box). These edge states are absent in the spectrum obtained from the periodic boundary conditions (pbc). We explicitly see the manifestation of the features associated to the non-trivial topology in the MM limit.

\subsection{Berry curvature from the slope of the Wannier charge center}
In this section we define the slope of the Wannier charge center to be proportional to the Berry curvature. In the case of quasiperiodic MM (incommensure modulation) the presence of an additional quasiperiodic translation symmetry which requires equivalence of translation along the site index $n$ with a translation in the space of the parameter $\varphi$, requires the independence of the Berry curvature on the parameter $\varphi$~Ref.~\cite{kraus1}. Below we show that this is reflected in the slope of Wannier charge center plot as a function of $\varphi$ being independent of $\varphi$. The spectral projector was defined as,
\[
\rho(n,\varphi)=\sum_{i\in occupied}\psi_{i}^{*}(n,\varphi)\psi_{i}(n,\varphi)
\]
The Wannier charge center can be written as, 
\begin{gather}
\bar{n}(\varphi)=\sum_{n}n\rho(n,\varphi) \\
\bar{n}(\varphi)=\sum_{n}\sum_{i\in occupied}n\psi_{i}^{*}(n,\varphi)\psi_{i}(n,\varphi)
\end{gather}
To calculate the slope we differentiate this with respect to $\varphi$ on both sides,
\bea
\frac{\partial\bar{n}(\varphi)}{\partial\varphi}&=&\sum_{n}\sum_{i\in occupied}n\bigg(\frac{\partial\psi_{i}^{*}(n,\varphi)}{\partial\varphi}\psi_{i}(n,\varphi)\nonumber\\
&+&\psi_{i}^{*}(n,\varphi)\frac{\partial\psi_{i}(n,\varphi)}{\partial\varphi}\bigg)\la{slope1}
\eea
We can write, 
\[
n\ \psi_{i}(n,\varphi)=\int\frac{d\theta}{2\pi i}\psi_{i}(\theta,\varphi)\frac{d}{d\theta}(e^{i\theta n})
\]
Substituting the above expression in Eq.~(\ref{slope1}) and integrating by parts we get,
\be
\frac{\partial\bar{n}(\varphi)}{\partial\varphi}=\sum_{i\in occupied}\sum_{n}e^{in\theta}\int\frac{d\theta}{2\pi i}\text{Im}\left(\frac{\partial\psi_{i}^{*}(n,\varphi)}{\partial\varphi}\frac{\partial\psi_{i}(\theta,\varphi)}{\partial\theta}\right)\nonumber\\
\ee
We can identify that the sum over $n$ can be absorbed as a fourier transform from $n$ to $\theta$. The resulting expression is the Berry curvature integrated over $\theta$ parameter. 
\be
\frac{\partial\bar{n}(\varphi)}{\partial\varphi}=\sum_{i\in occupied}\int\frac{d\theta}{2\pi i}\text{Im}[(\partial_{\varphi}\psi_{i}\partial_{\theta}\psi_{i})]\la{slope2}
\ee
Note that the $\theta$ integral is redundant too as the berry curvature is a property of the bulk bands and is independent of the twisted boundary conditions that $\theta$ imposes. We can directly see that for a constant slope the berry curvature us quantized and the Chern number is completely determined by the right hand side of Eq.~(\ref{slope2}).

\bibliography{references.bib}

\end{document}